\documentclass[aps, preprint, superscriptaddress, longbibliography]{revtex4-2}

\usepackage{graphicx,epsfig} 
\usepackage{amsmath}
\usepackage{amsfonts}
\usepackage{cancel}
\usepackage{bm}
\usepackage{amssymb}
\usepackage{amsmath,lipsum}
\usepackage[colorlinks=true, allcolors=blue]{hyperref}
\usepackage{soul}
\usepackage{etoolbox} 
\usepackage{orcidlink}

\newcommand{\ve}[1]{\boldsymbol{#1}}

\begin{document}
\title{
Generating ferro-spinetic polarizations in altermagnetic insulators
}

\author{Toshihiro Sato}
\affiliation{{Institute for Theoretical Solid State Physics, IFW Dresden, 01069 Dresden, Germany}}
\affiliation{{W\"urzburg-Dresden Cluster of Excellence ct.qmat, Germany}}

\author{Mengli Hu}
\affiliation{{Institute for Theoretical Solid State Physics, IFW Dresden, 01069 Dresden, Germany}}

\author{Ion Cosma Fulga}
\affiliation{{Institute for Theoretical Solid State Physics, IFW Dresden, 01069 Dresden, Germany}}
\affiliation{{W\"urzburg-Dresden Cluster of Excellence ct.qmat, Germany}}

\author{Oleg Janson}
\affiliation{{Institute for Theoretical Solid State Physics, IFW Dresden, 01069 Dresden, Germany}}

\author{Jorge I. Facio}
\affiliation{{Centro At\'{o}mico Bariloche, Instituto de Nanociencia y Nanotecnolog\'{\i}a and Instituto Balseiro, 9500 Av. Bustillo, Argentina}}

\author{Alessandro Stroppa}
\affiliation{{CNR-SPIN, Department of Physical and Chemical Sciences, University of L'Aquila, 67100 L'Aquila, Italy
}}

\author{Fakher F. Assaad\,\orcidlink{0000-0002-3302-9243}}
\affiliation{{Institut f\"ur Theoretische Physik und Astrophysik, Universit\"at W\"urzburg, 97074 W\"urzburg, Germany}}
\affiliation{{W\"urzburg-Dresden Cluster of Excellence ct.qmat, Germany}}

\author{Jeroen van den Brink}
\affiliation{{Institute for Theoretical Solid State Physics, IFW Dresden, 01069 Dresden, Germany}}
\affiliation{{W\"urzburg-Dresden Cluster of Excellence ct.qmat, Germany}}

\date{\today}

\begin{abstract}
Altermagnets are a novel class of fully spin-compensated magnetic materials that nevertheless have spin-split electronic bands, offering novel perspectives for spintronics applications.
Based on a rigorous analysis of altermagnetic many-body models and their symmetry we establish the important role of two fundamental types of polarizations in altermagnetic insulators: the charge and the spinetic one, where the latter corresponds to a macroscopic spin-displacement field. 
First principles calculations confirm and quantify their presence in real materials. 
The two polarizations are directly coupled and emerge in orthogonal directions when inversion symmetry is broken, either by the system developing a spontaneously ferroelectric polarization (in ferroelectric altermagnets), or by a charge displacement induced by an external electric field (for inversion invariant altermagnetic insulators). 
This presence of large and switchable spin accumulations constitute an attractive fundamental feature of altermagnetic insulators.
\end{abstract}

\maketitle

Ferroelectric insulators famously exhibit a spontaneous electric polarization that can be reversed by external electric fields.  Their electric polarization is caused by a net displacement of the valence band electron density with respect to the counter-charged atomic cores. The rich electrical, mechanical, and thermal properties of ferroelectrics find applications in a wide range of devices for data storage, sensing, and opto-electronics~\cite{Sekine2022}. From the observation that ferroelectric polarization induces charge accumulation of opposite sign at opposing surfaces of a material, it is clear that ferroelectric ordering breaks spatial inversion symmetry.
Going beyond non-magnetic ferroelectrics, we explore here the consequences of inversion symmetry breaking for altermagnets (A$\ell$Ms). A$\ell$Ms constitute a novel, third fundamental class of collinear magnetic ordered materials, alongside with ferro- and antiferromagnets. They share with conventional antiferromagnets the feature of a vanishing net magnetization, enforced by the A$\ell$M symmetry. At the same time they exhibit a spin-splitting of electronic valence and conduction bands, just as in ferromagnets~\cite{Smejkal20,Smejkal22,Smejkal22a,Naka19,Yuan20,Naka21,Guo23}.
The spin-splitting gives rise to various interesting phenomena such as an anomalous Hall effect, large Edelstein responses, topological surface states and magnon chiralities~\cite{Libor20, Libor21, RuO-22, Betancourt23, Bai22, Guo23_1, AM-obs, SatoT24_1,Li_24_1,Hu_24_1, Trama_24_1,Yershov24_1,Kravchuk_25_1}.

\begin{figure}
\centering
\centerline{\includegraphics[width=0.5\textwidth]{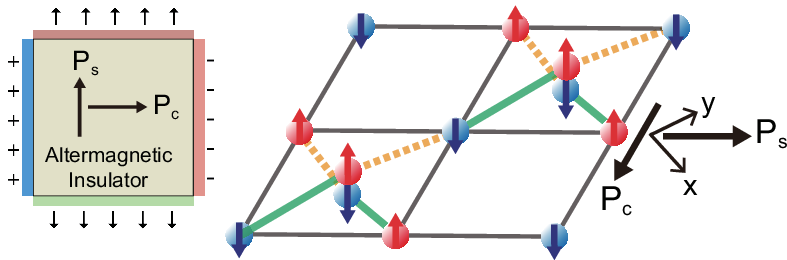}}
\caption{\label{fig:schematic} Left: inversion symmetry breaking inducing orthogonal ferroelectric polarization $\mathbf{P}_c$ and ferro-spinetic polarization $\mathbf{P}_s$ in a $d$-wave altermagnetic insulator. Right: 
Schematic of the hopping structure interacting fermionic lattice model, its emerging altermagnetic order and ferro-spinetic/ferroelectric polarizations. }
\end{figure}

Here we show that the spin-splitting of A$\ell$M bands generates a non-trivial response to spatial inversion symmetry breaking, particularly in the \emph{spin} sector. In charge sector, instead, the A$\ell$M response is similar to that of non-magnetic insulators, giving rise to a finite charge displacement $\mathbf{P}_c$. Due to the A$\ell$M spin-splitting apart from $\mathbf{P}_c$, also a spin displacement $\mathbf{P}_s$ is generated (see Fig.~\ref{fig:schematic}). This spin displacement, which we define precisely in the following and refer to as \emph{ferro-spinetic} polarization, has several remarkable properties. First of all it causes a net \emph{spin} accumulation at the surfaces, while at the same time due to its A$\ell$M symmetry the system does \emph{not} develop a net magnetization and remains fully spin-compensated. For $d$-wave A$\ell$Ms these polarizations are in orthogonal directions: $\mathbf{P}_c\perp \mathbf{P}_s$. Thus in A$\ell$Ms the ferro-spinetic and charge polarization are directly coupled, and proportional to each other. Indeed the direction the magnetic moments stemming from the spin density accumulated at the surface is switched upon inverting the charge dipole $\mathbf{P}_c$. In practice this may be achieved by applying an electric field $\mathbf{E}$ to the material. Interestingly, the same ferro-spinetic polarization also emerges in A$\ell$Ms with inversion symmetric lattices, in which the charge polarization is induced  by an external electric field. As long as the direction of $\mathbf{E}$ preserves A$\ell$M symmetry, it generates a ferro-spinetic polarization $\mathbf{P}_s$ in the orthogonal direction. The switchability of $\mathbf{P}_s$ by an orthogonal electric field suggests a pathway to A$\ell$M spintronics applications.

In the following we first demonstrate the presence of ferro-spinetic polarization $\mathbf{P}_s$ in its purest form in an interacting altermagnetic fermion model, in which a many-body chiral symmetry explicitly forbids the presence of charge polarization. Consequently inversion symmetry breaking induces only ferro-spinetic polarization, which is evidenced by our quantum Monte Carlo simulations revealing the presence of reversible, edge-localized spin accumulations. Breaking the chiral symmetry of the model then releases the charge sector, which subsequently yields a ferroelectric polarization $\mathbf{P}_c$ orthogonal to $\mathbf{P}_s$. We then identify existing Mn-based metal-organic frameworks (MOFs) as hosts for this effect. Calculating the ferroelectric and ferro-spinetic polarizations of a Mn-MOF from first principles using the modern theory of polarization, we find a ferro-spinetic polarization in this material that is even larger than its ferroelectric one.
We should note that the ferro-spinetic effect is very different from multiferroicity: due to the A$\ell$M symmetry the net magnetization vanishes at all times.

\section*{Definition of $\mathbf{P}_c$ and $\mathbf{P}_s$}
Before introducing the models and real materials, it is advantageous to formally define the ferroelectric/ferro-spinetic polarizations $\mathbf{P}_c$ and $\mathbf{P}_s$.  In a finite system the electric dipole moment $\mathbf{P}_c$ simply corresponds to the first moment of the charge density $\rho(\mathbf{r})$ (depending on position $\mathbf{r}$) via the volume integral $ \int_V \mathbf{r} \rho_c (\mathbf{r}) d^3\mathbf{r}$. In the presence of spinful electrons, $\rho_c$ is the sum of the spin sectors $\rho_c=\rho_\uparrow+\rho_\downarrow$, where for convenience we suppressed the contribution of the atomic cores. This separation into two spin sectors is appropriate in absence of relativistic spin-orbit coupling, which is the relevant A$\ell$M limit~\footnote{Generalization to the case with finite spin-orbit coupling is straightforward.}. The modern theory of polarization~\cite{King-Smith1993,Resta1994,Resta1998} now allows evaluation of charge displacement of an infinite, periodic solid from the Bloch states $|u^\sigma_{n \mathbf{k}}\rangle$, where $\sigma$ and $n$ are spin and band index, respectively, and $\mathbf{k}$ crystal momentum. Defining the Berry connection $\mathbf{A}^\sigma_n (\mathbf{k}) = i \langle u^\sigma_{n \mathbf{k}} |\partial_\mathbf{k}| u^\sigma_{n \mathbf{k}} \rangle$, one obtains for the polarization in a spin sector  
$\mathbf{P}_\sigma = \frac{i e}{(2\pi)^3} \sum_{n} \int_{BZ}  \mathbf{A}^\sigma_n (\mathbf{k}) \ d^3 \mathbf{k}$, so that the charge displacement is the sum over the two spin sectors: $\mathbf{P}_c =  \mathbf{P}_\uparrow+\mathbf{P}_\downarrow$. 
In this setting it is natural to define also a spin displacement field as  $\mathbf{P}_s =  \mathbf{P}_\uparrow - \mathbf{P}_\downarrow$, here referred to as ferro-spinetic polarization~\footnote{Ref.~\onlinecite{Saez2025} coins the term ferro-spintronic order for the dipole moment of the spin density. As spintronics is an already established field of technology, we prefer not to conflate our nomenclature with that terminology.}. 
Equivalently, in a finite system $\mathbf{P}_s$ corresponds to the first moment of the spin density distribution $\rho_s=\rho_\uparrow-\rho_\downarrow$. The time derivative of $\mathbf{P}_s$ then defines a proper and measurable spin-current~\cite{Shi2006}.

\section*{Interacting altermagnetic model}
To determine the effect of inversion symmetry breaking in a proper many-body setting, we define a 2D Hubbard model Hamiltonian of appropriate symmetry. In its A$\ell$M Mott-insulating regime the model is tractable by quantum Monte Carlo simulations, allowing determination of the relevant polarizations in an unbiased manner. The Hamiltonian
\begin{eqnarray}
\hat{H}=-\sum_{{\langle \ve{i}, \ve{j} \rangle},s} t_{\ve{i},\ve{j}} \hat{c}_{\ve{i}s}^{\dagger} \hat{c}^{}_{\ve{j}s} 
+U\sum_{\ve{i}} \left(\hat{n}_{\ve{i}\uparrow}-\frac{1}{2} \right)\left(\hat{n}_{\ve{i}\downarrow}-\frac{1}{2} \right)
\label{model}
\end{eqnarray}
consists of a checkerboard lattice with a four-site unit cell, whose sites are denoted by A, B, C, and D, arranged as shown schematically in Fig.~\ref{fig:schematic}.
Here, $\hat{c}_{\ve{i}s}^{\dagger} $  creates a fermion with spin $s = \uparrow, \downarrow$ at site $\ve{i}$.
Sites A and B are nearest neighbors on the underlying square lattice and are connected by a nearest-neighbor hopping $t$.
Sites C and D are located at the centers of alternating plaquettes of the square lattice.
Along the $+y$ diagonal direction, the hopping between sites A and C alternates between $t'$ and $t''$ from plaquette to plaquette.
Along the orthogonal diagonal ($+x$) direction, the hopping between sites B and D alternates, with $t'$ and $t''$ interchanged.
These diagonal hoppings are parametrized as $t' = t_1+\delta t_1/2$ and $t'' = t_1-\delta t_1/2$.
The repulsive onsite Hubbard interaction ($U>0$) with $\hat{n}_{\ve{i}s} \equiv \hat{c}_{\ve{i}s}^\dagger \hat{c}^{\phantom{\dagger}}_{\ve{i}s}$ drives long-range magnetic ordering.
The presence of sites C and D avoids geometric frustration and favors fully compensated collinear magnetic order that \emph{cannot} be mapped onto itself by a combination of time-reversal and a lattice translation or inversion, thereby realizing an altermagnet. $\hat{H}$ is invariant under a particle-hole transformation so that our choice of chemical potential corresponds to half-filling ($\hat{n}_{\ve{i}s}=1/2$). A finite value of $\delta t_1$ breaks the $\mathcal{C}_{4}$ rotation symmetry, inversion symmetry, and the glide symmetry $\mathcal{G}_{x+y}$, whereas the SU(2) spin symmetry, time-reversal symmetry $\mathcal{T}$, and the orthogonal glide $\mathcal{G}_{x-y}$ remain unbroken. Importantly, $\hat H$ possesses a many-body chiral symmetry $\hat U_\Gamma$ that remains unbroken also after the A$\ell$M order emerges. The explicit form and derivation of $\hat U_\Gamma$ are detailed in the Supplemental Information. 
The unbroken chiral symmetry pins the ferroelectric polarization to zero while still allowing for a ferro-spinetic polarization, as introduced above.

\section*{Emerging A$\ell$M ordering}
To establish the presence of magnetic order in the Mott-insulating phase, we employ unbiased quantum Monte Carlo simulations~\cite{ALF_v1, ALF_v2}.
The simulations allow us to access the ground-state properties of the interacting system, with details provided in the Supplemental Information.
We find that the system hosts an A$\ell$M Mott-insulating phase (see Fig.~\ref{fig:schematic}) for a wide range of parameters $U$ and $\delta t_1$.
A key feature of the altermagnetic insulating state is that for finite $\delta t_1$ both time-reversal $\mathcal{T}$ and glide $\mathcal{G}_{x-y}$ symmetry are spontaneously broken, while  their combination, $\mathcal{G}_{x-y}\mathcal{T}$, is preserved. This is the A$\ell$M symmetry that guarantees perfect spin compensation. When $|\delta t_1|$ becomes very large the system transitions into a valence-bond solid state.

\begin{figure}[t]
\centering
\centerline{\includegraphics[width=0.5\textwidth]{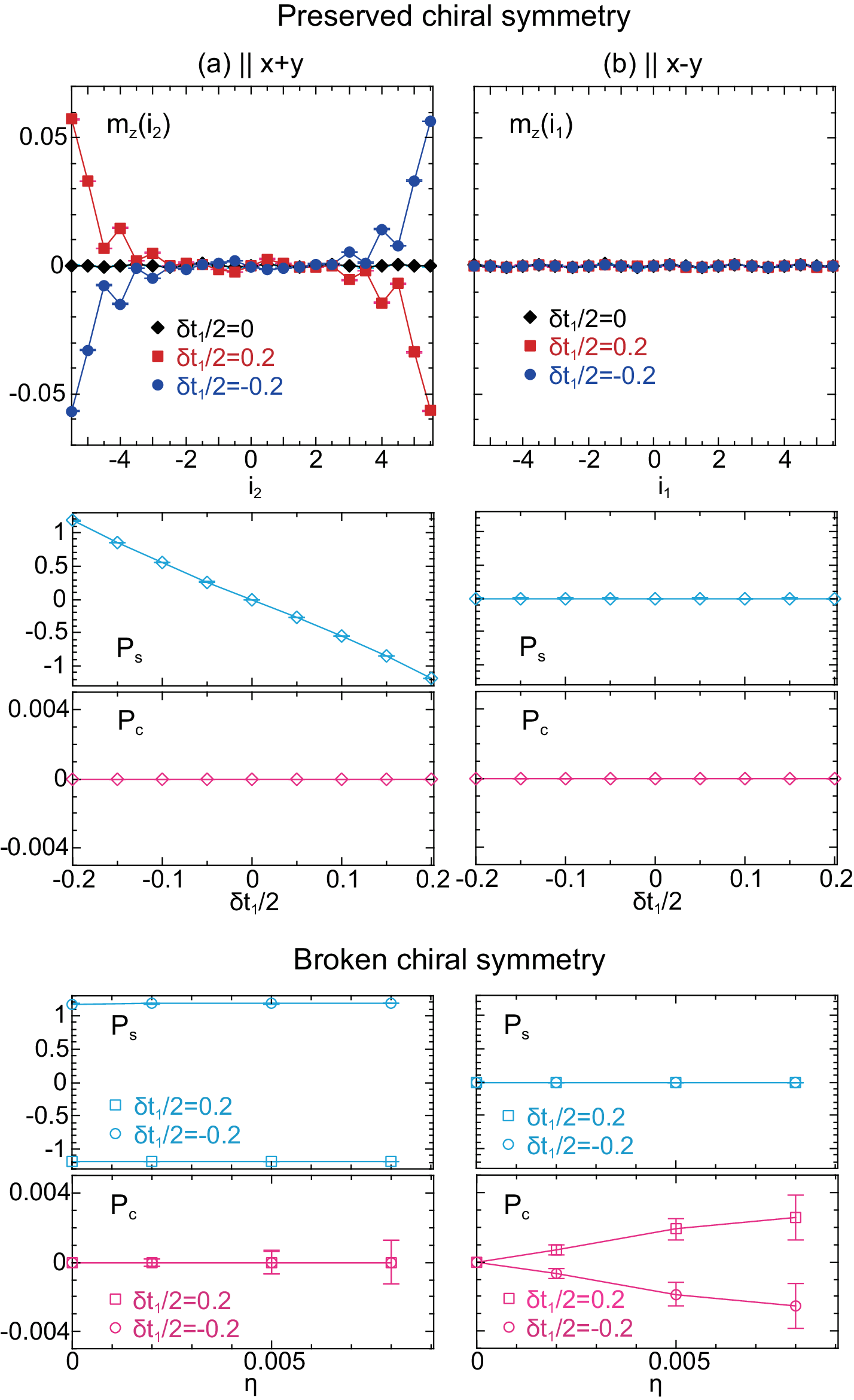}}
\caption{\label{fig:Polarization}
Real-space distribution of magnetization $m_z(i)$, from which the spin polarization is evaluated as $P_s=\sum_{i} i m^z(i)$. 
Panel (a) shows results for the $\parallel x+y$ geometry, with periodic boundaries along the $\mathbf{a}_2 - \mathbf{a}_1$ direction and open boundaries along the $\mathbf{a}_2 + \mathbf{a}_1$ direction. 
Panel (b) corresponds to the $\parallel x-y$ geometry, where the boundary conditions are reversed.
These results correspond to a case where chiral symmetry is preserved.
A finite $\delta t_1$ breaks inversion symmetry.
The charge polarization $P_c$ is similarly evaluated by replacing $m_z(i)$ with the charge distribution.
Bottom panels: polarization quantities $P_s$ and $P_c$, calculated under the same conditions and geometry as in the top panels, but in a case where chiral symmetry is broken. 
A finite $\eta$ breaks chiral symmetry.
}
\end{figure}

\section*{Ferroelectric and ferro-spinetic polarizations}
We now show that the A$\ell$M ground state produces a ferro-spinetic polarization upon inversion symmetry breaking. As we are dealing with a finite system, we can evaluate $P_s$ directly from the first moment of the spin-distribution, employing open boundary conditions in one direction and periodic in the other. A spin accumulation at the open boundaries that is independent of termination then evidences the presence of a ferro-spinetic polarization. We first consider systems with chiral symmetry, where $P_c$ must vanish, and we are dealing with a purely ferro-spinetic polarization. 
Thus, we employ a cylindrical geometry with periodic boundary conditions along the $x$-$y$ direction and open boundary conditions along $x$+$y$.
We label the spatial coordinate along the $x$+$y$ direction by $i_1$, while $i_2$ denotes the position along the $x$-$y$ direction.
Details of the numerical setup are provided in the Supplemental Information.

To determine the profile of the spin distribution, we measure the real-space distribution of magnetization in the direction for which we have open boundary conditions by evaluating for each row the magnetizations in the periodic direction $m_z(i_2)=1/L\sum_{i_1}\langle \hat{S}^z_{i_1,i_2}\rangle$.
Figure~\ref{fig:Polarization}(a) shows this quantity in the altermagnetic insulating phase.
For all cases considered there is no net magnetization. 
However, the key question is whether inversion breaking induces a spin accumulation at the edges of the system. 
For $\delta t_1 = 0$ the system has inversion symmetry and we observe that the profile of $m_z(i_2)$ is flat, indicating the absence of spin accumulation.
For finite $\delta t_1$ there is a clear magnetization imbalance between the regions near the two edges compared to the bulk, leading to an appreciable accumulation of spin. This imbalance does not depend on the edge termination.
Specifically, a negative $m_z$ is observed at one edge and a positive $m_z$ at the other, with the accumulation extending over several lattice spacings. 
We observe that reversing the inversion symmetry breaking (by reversing the sign of $\delta t_1$) switches the sign of the spin accumulation while leaving its spatial profile unaltered.

Direct evaluation of the first moment of the spin distribution in the $x$+$y$ direction, $P_s=\sum_{i_2} i_2 m^z(i_2)$, first of all confirms that for the inversion symmetric ($\delta t_1$=0) case  $P_s$ vanishes, see Fig.~\ref{fig:Polarization} -- indeed $P_s$ is observed to be proportional to $\delta t_1$. 
Scaling analysis indicates that in the thermodynamic limit $P_s$ converges to a finite value (see Supplemental Information). 
Along the orthogonal $x$-$y$ direction $P_s$ vanishes.
This anisotropy is due to the symmetry of the altermagnetic state.  When the N\'eel order sets in, both time-reversal $\mathcal T$ and glide $\mathcal G_{x-y}$ symmetry are broken individually, but their product $\mathcal{G}_{x-y}\mathcal {T}$ remains intact. With respect to this combined symmetry, the $x$-$y$ component of $P_s$ is odd and hence forbidden, whereas the $x$+$y$ component is even and thus allowed, locking the ferro-spinetic polarization to the $x$+$y$ axis.

\begin{figure}[t]
\centering
\centerline{\includegraphics[width=0.5\textwidth]{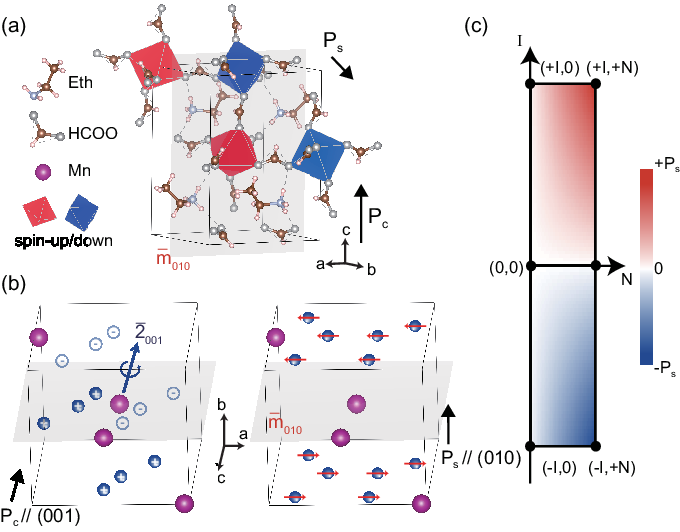}}
\caption{\label{fig:MnMOF}
Ferro-spinetic and altermagnetic properties of
[C$_2$H$_5$NH$_3$]Mn[(HCOO)$_3$] (Mn-MOF). (a) Crystal and magnetic structures. 
The building blocks of Mn-MOF are the ethylammonium: C$_2$H$_5$NH$_3^+$, divalent Mn$^{2+}$ ion, and the carboxylate HCOO$^-$. 
The spin ($P_s$) and charge ($P_c$) polarization directions are labelled in (b). 
With screw-rotation ($\overline{2}_{001}$) and glide-mirror symmetries ($\overline{m}_{010}$), $P_c$ and $P_s$ are aligned along $z$ and $y$ axes, respectively. 
(c) Spin polarization $P_s$, illustrating the effect of inversion symmetry ($I$). 
The sign of $P_s$ reverses when only the structural configuration is inversion inverted, while keeping the direction of the N\'eel order $(N)$ fixed, confirming the transformation property of $P_s$ under inversion symmetry. 
}
\end{figure}

\section*{Breaking Chiral Symmetry }
In real materials particle-hole symmetry is always broken to some extent. We therefore explicitly break the many-body chiral symmetry $\hat U_\Gamma$ in the altermagnetic insulating state. While this barely affects the shape and size of the ferro-spinetic polarization, we will show that it induces an additional ferroelectric polarization in the orthogonal direction. An alternating onsite potential in the system breaks chirality and particle-hole symmetry. In the  Hamiltonian this symmetry breaking term takes the form $\sum_{i,s} \eta_i \hat{n}_{i.s}$, where $\eta_i =+\eta (-\eta)$ for $i \in \text{A},\text{B} (\text{C},\text{D})$
{\footnote{{Including $\eta$ term breaks chiral symmetry and introduces a negative sign problem; in the $\eta$ range explored here it remains mild (e.g., at $\eta=0.008$ we find quantum Monte  Carlo average sign,  $\langle \text{sign} \rangle \approx0.606$)}}}.
This term explicitly breaks $\hat U_\Gamma$ while preserving the altermagnetic symmetry.
Figure~\ref{fig:Polarization} shows that in the $x$+$y$ direction  $P_s$  persists when the chiral symmetry is broken.
The charge polarization $P_c$ along $x$+$y$ remains zero but along $x$-$y$  we observe that non-zero $\eta$ gives rise to finite $P_c$ (while $P_s$ remains zero). 
Changing the direction of inversion symmetry breaking by inverting the sign of $\delta t_1$ changes the sign of $P_c$. Also this behavior follows from the altermagnetic symmetry: as the charge operator $n(i)$ is even under $\mathcal{T}$, $P_c$ transforms under $\mathcal{G}_{x-y} \mathcal{T}$ so that the $x$+$y$ component is odd and therefore forbidden, whereas the $x$-$y$ component is even and thus allowed.
Consequently, once the many-body chiral symmetry is broken, a ferroelectric polarization  develops only along the $x$-$y$ direction, orthogonal to the ferro-spinetic one along $x$+$y$.

\section*{Real materials: Mn-based metal-organic frameworks (MOFs)}
To investigate the concomitant ferroelectric and ferro-spinetic polarizations in a real material quantitatively from first principles, we focus on MOFs~\cite{Mn-MOF_numerical}, that, besides meeting the  symmetry requirements, provide a large space of compounds. As representative altermagnet we select [C$_2$H$_5$NH$_3$]Mn[(HCOO)$_3$] (Mn-MOF, see Fig.~\ref{fig:MnMOF}), which is a polar magnetic insulator~\cite{Mn-MOF_numerical} with a relatively large band gap of 3.59 eV.
Inversion symmetry is broken in its crystal structure with magnetic space group $Pn'a2_1'$, whereas the screw-rotation and glide-mirror symmetries serve as the altermagnetic symmetries connecting opposite spins.  In the language of spin space groups, these constraints can be written as a combination of a (glide) mirror in real space and an inversion in spin space: $\{-1||m_y|\bm{t}\}$, which is equivalent to $\mathcal{G}_{x-y} \mathcal{T}$. As we have already seen in the two-dimensional ferro-spinetic model, $P_s$ and $P_c$ must be perpendicular and parallel to the mirror plane. The first-principles calculations (see the band structure in Supplementary Information) confirm this observation and we find $P_s = P_{\uparrow} - P_{\downarrow} = (0,13.95,0)\mu C (\hbar/2e) /cm^2$ and $P_c = (0,0,-1.45)\mu C/cm^2$.  In Fig.~\ref{fig:MnMOF} (b), the local charge/spin polarizations are indicated schematically, and the final polarization directions are shown through the crystal-symmetry constraints. The calculations confirm that the spin polarization $P_s$ changes sign under inversion symmetry.  This behavior is explicitly demonstrated in Fig.~\ref{fig:MnMOF} (c), where the direction of the N\'eel order is kept fixed while only the structural configuration related by inversion is reversed, resulting in a sign flip of $P_s$. Consistently, $P_s=0$ is found along the $x$ and $y$ axes. Thus the spin displacement  $P_s$ can be switched by an external perturbation, in particular an applied electric field, while maintaining overall zero total spin polarization.

From symmetry it follows that just as well a spinetic polarization is induced in \emph{inversion symmetric} altermagnetic insulators by application of an orthogonal external electric field. In the ground state inversion symmetry obviously prohibits both $P_c$ and $P_s$, while altermagnetic band splittings are not affected for the identical spin Laue groups~\cite{Smejkal22}. As a straightforward example we may consider the Mn-MOF above in its inversion symmetric lattice configuration with the higher magnetic space group symmetry $Pn'ma'$. 
Quite a number of (inversion symmetric) $d$-wave altermagnetic materials meet this criterion, including LaCrO$_3$, YCrO$_3$, LaMnO$_3$, Mn$_2$SeO$_3$F$_2$, KMnF$_3$, SmFeO$_3$, TbFeO$_3$, NdFeO$_3$, TbCrO$_3$, Mn(N(CN)$_2$)$_2$, Sr$_4$Fe$_4$O$_{11}$, and La$_2$NiO$_4$~\cite{Guo23,KNIGHT2020155935_kmnf3}.
An external electric field may now break their inversion symmetry and lower it to $Pn'a2_1'$.
For electric fields along the $y/z$ axis, $\{-1||m_{z/y}|\bm{t}_{2/1}\}$ is preserved, guaranteeing both the altermagnetic nature and a non-zero charge displacement $P_c \parallel y/z$ and spinetic polarization $P_s \parallel z/y$. If instead the electric field is applied along the $x$ axis both $\{-1||m_{y}|\bm{t}_1\}$ and $\{-1||m_{z}|\bm{t}_2\}$ are preserved, so that $P_s$ vanishes, indicating how the polarity and direction of the external electric field switches the spinetic polarization.

\section*{Outlook }
From a rigorous analysis of altermagnetic many-body models and their symmetries we established the important role of the charge and spinetic polarization in altermagnetic insulators. 
In its purest form we have demonstrated this in an interacting fermion model with chiral symmetry: unbiased quantum Monte Carlo simulations reveal that inversion breaking induces a reversible spin accumulation at the system's edges. Once chirality is broken an additional ferroelectric polarization emerges.
The fact that chiral symmetry suppresses ferroelectric polarization but leaves ferro-spinetic polarization intact highlights that ferro-spinetic polarization is a more general manifestation of inversion symmetry breaking in altermagnetic insulators.
First principles calculations confirm and quantify their presence in real materials: altermagnetic Mn-based metal-organic frameworks can carry a very large ferro-spinetic polarization which exceeds the ferroelectric more than tenfold.
The two polarizations are directly coupled and emerge in orthogonal directions when inversion symmetry is broken, be it by spontaneously developing ferroelectric polarization, or by a charge displacement induced by an external electric field. This response to an electric field with switchable magnetic surface polarizations allows altermagnetic insulators to act as charge-to-spin converters, which due to the electronic nature of both polarizations, may operate with minimal losses and at high speeds.

\section*{Methods}
We employ the ALF (Algorithms for Lattice Fermions) implementation~\cite{ALF_v1, ALF_v2} of the grand-canonical, finite-temperature auxiliary-field quantum Monte Carlo method~\cite{Blankenbecler81, White89, Assaad08_rev} to simulate the interacting altermagnetic model.
At half filling the model is free of the negative-sign problem.
Energies are measured in units of $t$, and we fix $t_1=0.8$.
All data were calculated for an inverse temperature $\beta=80$ (with Trotter discretization $\Delta\tau=0.1$).
In the considered parameter range this choice of temperature was sufficient to obtain results representative of the ground state.
Bulk properties are obtained using torus geometries with periodic boundary conditions in both directions on lattices of $L\times L$ unit cells ($4L^2$ sites), while ferro-spinetic polarizations are evaluated using cylindrical geometries with periodic boundary conditions along one direction and open boundary conditions along the orthogonal direction.
We simulate cylindrical lattices with $L=6$ unit cells along the periodic direction and $N_{\rm orb}=46$ orbitals along the open direction, fixing $U=5$.
In addition to the lattice-model calculations, we perform first-principles electronic-structure calculations for Mn-based metal-organic frameworks (MOFs), using the VASP code \cite{GKresse_PRB1996_JFurthmuller} with the projector augmented-wave method \cite{PEBlochl_PRB1994}.
The Brillouin zone was sampled using a $5\times5\times5$ $\Gamma$-centered $k$-point mesh.
The energy cutoff for the plane-wave basis was set to 600 eV.
The Hubbard term was introduced and set to 3.0 eV in the $d$ orbitals of the Mn atom in the DFT framework (DFT+U) to account for electron-electron correlations.

\section*{Acknowledgments}
We thank Rembert Duine for fruitful discussions. We gratefully acknowledge the Gauss Centre for Supercomputing e.V. for funding this project by providing computing time on the GCS Supercomputer SUPERMUC-NG at Leibniz Supercomputing Centre, (project No.~pn73xu)
as well as the scientific support and HPC resources provided by the Erlangen National High Performance Computing Center (NHR@FAU) of the Friedrich-Alexander-Universit\"at Erlangen-N\"urnberg (FAU) under the NHR project b133ae. 
NHR funding is provided by federal and Bavarian state authorities. 
NHR@FAU hardware is partially funded by the German Research Foundation (DFG) --- 440719683.
FA, TS, JvdB and ICF thank the W\"urzburg-Dresden Cluster of Excellence on Complexity and Topology in Quantum Matter ct.qmat (EXC 2147, project-id 390858490). 
MH thanks the Leibniz Association through Leibniz Competition Project No. J200/2024.


\begin{thebibliography}{48}%
\makeatletter
\providecommand \@ifxundefined [1]{%
 \@ifx{#1\undefined}
}%
\providecommand \@ifnum [1]{%
 \ifnum #1\expandafter \@firstoftwo
 \else \expandafter \@secondoftwo
 \fi
}%
\providecommand \@ifx [1]{%
 \ifx #1\expandafter \@firstoftwo
 \else \expandafter \@secondoftwo
 \fi
}%
\providecommand \natexlab [1]{#1}%
\providecommand \enquote  [1]{``#1''}%
\providecommand \bibnamefont  [1]{#1}%
\providecommand \bibfnamefont [1]{#1}%
\providecommand \citenamefont [1]{#1}%
\providecommand \href@noop [0]{\@secondoftwo}%
\providecommand \href [0]{\begingroup \@sanitize@url \@href}%
\providecommand \@href[1]{\@@startlink{#1}\@@href}%
\providecommand \@@href[1]{\endgroup#1\@@endlink}%
\providecommand \@sanitize@url [0]{\catcode `\\12\catcode `\$12\catcode
  `\&12\catcode `\#12\catcode `\^12\catcode `\_12\catcode `\%12\relax}%
\providecommand \@@startlink[1]{}%
\providecommand \@@endlink[0]{}%
\providecommand \url  [0]{\begingroup\@sanitize@url \@url }%
\providecommand \@url [1]{\endgroup\@href {#1}{\urlprefix }}%
\providecommand \urlprefix  [0]{URL }%
\providecommand \Eprint [0]{\href }%
\providecommand \doibase [0]{https://doi.org/}%
\providecommand \selectlanguage [0]{\@gobble}%
\providecommand \bibinfo  [0]{\@secondoftwo}%
\providecommand \bibfield  [0]{\@secondoftwo}%
\providecommand \translation [1]{[#1]}%
\providecommand \BibitemOpen [0]{}%
\providecommand \bibitemStop [0]{}%
\providecommand \bibitemNoStop [0]{.\EOS\space}%
\providecommand \EOS [0]{\spacefactor3000\relax}%
\providecommand \BibitemShut  [1]{\csname bibitem#1\endcsname}%
\let\auto@bib@innerbib\@empty
\bibitem [{\citenamefont {Sekine}\ \emph {et~al.}(2022)\citenamefont {Sekine},
  \citenamefont {Akiyoshi},\ and\ \citenamefont {Hayami}}]{Sekine2022}%
  \BibitemOpen
  \bibfield  {author} {\bibinfo {author} {\bibfnamefont {Y.}~\bibnamefont
  {Sekine}}, \bibinfo {author} {\bibfnamefont {R.}~\bibnamefont {Akiyoshi}},\
  and\ \bibinfo {author} {\bibfnamefont {S.}~\bibnamefont {Hayami}},\
  }\bibfield  {title} {\bibinfo {title} {Recent advances in ferroelectric metal
  complexes},\ }\href
  {https://doi.org/https://doi.org/10.1016/j.ccr.2022.214663} {\bibfield
  {journal} {\bibinfo  {journal} {Coordination Chemistry Reviews}\ }\textbf
  {\bibinfo {volume} {469}},\ \bibinfo {pages} {214663} (\bibinfo {year}
  {2022})}\BibitemShut {NoStop}%
\bibitem [{\citenamefont {{\v{S}}mejkal}\ \emph {et~al.}(2020)\citenamefont
  {{\v{S}}mejkal}, \citenamefont {Gonz{\'{a}}lez-Hern{\'{a}}ndez},
  \citenamefont {Jungwirth},\ and\ \citenamefont {Sinova}}]{Smejkal20}%
  \BibitemOpen
  \bibfield  {author} {\bibinfo {author} {\bibfnamefont {L.}~\bibnamefont
  {{\v{S}}mejkal}}, \bibinfo {author} {\bibfnamefont {R.}~\bibnamefont
  {Gonz{\'{a}}lez-Hern{\'{a}}ndez}}, \bibinfo {author} {\bibfnamefont
  {T.}~\bibnamefont {Jungwirth}},\ and\ \bibinfo {author} {\bibfnamefont
  {J.}~\bibnamefont {Sinova}},\ }\bibfield  {title} {\bibinfo {title} {Crystal
  time-reversal symmetry breaking and spontaneous hall effect in collinear
  antiferromagnets},\ }\href {https://doi.org/10.1126/sciadv.aaz8809}
  {\bibfield  {journal} {\bibinfo  {journal} {Science Advances}\ }\textbf
  {\bibinfo {volume} {6}},\ \bibinfo {pages} {eaaz8809} (\bibinfo {year}
  {2020})}\BibitemShut {NoStop}%
\bibitem [{\citenamefont {\ifmmode~\check{S}\else \v{S}\fi{}mejkal}\ \emph
  {et~al.}(2022)\citenamefont {\ifmmode~\check{S}\else \v{S}\fi{}mejkal},
  \citenamefont {Sinova},\ and\ \citenamefont {Jungwirth}}]{Smejkal22}%
  \BibitemOpen
  \bibfield  {author} {\bibinfo {author} {\bibfnamefont {L.}~\bibnamefont
  {\ifmmode~\check{S}\else \v{S}\fi{}mejkal}}, \bibinfo {author} {\bibfnamefont
  {J.}~\bibnamefont {Sinova}},\ and\ \bibinfo {author} {\bibfnamefont
  {T.}~\bibnamefont {Jungwirth}},\ }\bibfield  {title} {\bibinfo {title}
  {Beyond conventional ferromagnetism and antiferromagnetism: A phase with
  nonrelativistic spin and crystal rotation symmetry},\ }\href
  {https://doi.org/10.1103/PhysRevX.12.031042} {\bibfield  {journal} {\bibinfo
  {journal} {Phys. Rev. X}\ }\textbf {\bibinfo {volume} {12}},\ \bibinfo
  {pages} {031042} (\bibinfo {year} {2022})}\BibitemShut {NoStop}%
\bibitem [{\citenamefont {{\v{S}}mejkal}\ \emph {et~al.}(2022)\citenamefont
  {{\v{S}}mejkal}, \citenamefont {Sinova},\ and\ \citenamefont
  {Jungwirth}}]{Smejkal22a}%
  \BibitemOpen
  \bibfield  {author} {\bibinfo {author} {\bibfnamefont {L.}~\bibnamefont
  {{\v{S}}mejkal}}, \bibinfo {author} {\bibfnamefont {J.}~\bibnamefont
  {Sinova}},\ and\ \bibinfo {author} {\bibfnamefont {T.}~\bibnamefont
  {Jungwirth}},\ }\bibfield  {title} {\bibinfo {title} {Emerging research
  landscape of altermagnetism},\ }\href
  {https://doi.org/10.1103/physrevx.12.040501} {\bibfield  {journal} {\bibinfo
  {journal} {Physical Review X}\ }\textbf {\bibinfo {volume} {12}},\ \bibinfo
  {pages} {040501} (\bibinfo {year} {2022})}\BibitemShut {NoStop}%
\bibitem [{\citenamefont {{Naka}}\ \emph {et~al.}(2019)\citenamefont {{Naka}},
  \citenamefont {{Hayami}}, \citenamefont {{Kusunose}}, \citenamefont
  {{Yanagi}}, \citenamefont {{Motome}},\ and\ \citenamefont {{Seo}}}]{Naka19}%
  \BibitemOpen
  \bibfield  {author} {\bibinfo {author} {\bibfnamefont {M.}~\bibnamefont
  {{Naka}}}, \bibinfo {author} {\bibfnamefont {S.}~\bibnamefont {{Hayami}}},
  \bibinfo {author} {\bibfnamefont {H.}~\bibnamefont {{Kusunose}}}, \bibinfo
  {author} {\bibfnamefont {Y.}~\bibnamefont {{Yanagi}}}, \bibinfo {author}
  {\bibfnamefont {Y.}~\bibnamefont {{Motome}}},\ and\ \bibinfo {author}
  {\bibfnamefont {H.}~\bibnamefont {{Seo}}},\ }\bibfield  {title} {\bibinfo
  {title} {{Spin current generation in organic antiferromagnets}},\ }\href
  {https://doi.org/10.1038/s41467-019-12229-y} {\bibfield  {journal} {\bibinfo
  {journal} {Nature Communications}\ }\textbf {\bibinfo {volume} {10}},\
  \bibinfo {eid} {4305} (\bibinfo {year} {2019})}\BibitemShut {NoStop}%
\bibitem [{\citenamefont {Yuan}\ \emph {et~al.}(2020)\citenamefont {Yuan},
  \citenamefont {Wang}, \citenamefont {Luo}, \citenamefont {Rashba},\ and\
  \citenamefont {Zunger}}]{Yuan20}%
  \BibitemOpen
  \bibfield  {author} {\bibinfo {author} {\bibfnamefont {L.-D.}\ \bibnamefont
  {Yuan}}, \bibinfo {author} {\bibfnamefont {Z.}~\bibnamefont {Wang}}, \bibinfo
  {author} {\bibfnamefont {J.-W.}\ \bibnamefont {Luo}}, \bibinfo {author}
  {\bibfnamefont {E.~I.}\ \bibnamefont {Rashba}},\ and\ \bibinfo {author}
  {\bibfnamefont {A.}~\bibnamefont {Zunger}},\ }\bibfield  {title} {\bibinfo
  {title} {Giant momentum-dependent spin splitting in centrosymmetric low-$z$
  antiferromagnets},\ }\href {https://doi.org/10.1103/PhysRevB.102.014422}
  {\bibfield  {journal} {\bibinfo  {journal} {Phys. Rev. B}\ }\textbf {\bibinfo
  {volume} {102}},\ \bibinfo {pages} {014422} (\bibinfo {year}
  {2020})}\BibitemShut {NoStop}%
\bibitem [{\citenamefont {Naka}\ \emph {et~al.}(2021)\citenamefont {Naka},
  \citenamefont {Motome},\ and\ \citenamefont {Seo}}]{Naka21}%
  \BibitemOpen
  \bibfield  {author} {\bibinfo {author} {\bibfnamefont {M.}~\bibnamefont
  {Naka}}, \bibinfo {author} {\bibfnamefont {Y.}~\bibnamefont {Motome}},\ and\
  \bibinfo {author} {\bibfnamefont {H.}~\bibnamefont {Seo}},\ }\bibfield
  {title} {\bibinfo {title} {Perovskite as a spin current generator},\ }\href
  {https://doi.org/10.1103/PhysRevB.103.125114} {\bibfield  {journal} {\bibinfo
   {journal} {Phys. Rev. B}\ }\textbf {\bibinfo {volume} {103}},\ \bibinfo
  {pages} {125114} (\bibinfo {year} {2021})}\BibitemShut {NoStop}%
\bibitem [{\citenamefont {Guo}\ \emph {et~al.}(2023{\natexlab{a}})\citenamefont
  {Guo}, \citenamefont {Liu}, \citenamefont {Janson}, \citenamefont {Fulga},
  \citenamefont {{van den Brink}},\ and\ \citenamefont {Facio}}]{Guo23}%
  \BibitemOpen
  \bibfield  {author} {\bibinfo {author} {\bibfnamefont {Y.}~\bibnamefont
  {Guo}}, \bibinfo {author} {\bibfnamefont {H.}~\bibnamefont {Liu}}, \bibinfo
  {author} {\bibfnamefont {O.}~\bibnamefont {Janson}}, \bibinfo {author}
  {\bibfnamefont {I.~C.}\ \bibnamefont {Fulga}}, \bibinfo {author}
  {\bibfnamefont {J.}~\bibnamefont {{van den Brink}}},\ and\ \bibinfo {author}
  {\bibfnamefont {J.~I.}\ \bibnamefont {Facio}},\ }\bibfield  {title} {\bibinfo
  {title} {Spin-split collinear antiferromagnets: A large-scale ab-initio
  study},\ }\href
  {https://doi.org/https://doi.org/10.1016/j.mtphys.2023.100991} {\bibfield
  {journal} {\bibinfo  {journal} {Materials Today Physics}\ }\textbf {\bibinfo
  {volume} {32}},\ \bibinfo {pages} {100991} (\bibinfo {year}
  {2023}{\natexlab{a}})}\BibitemShut {NoStop}%
\bibitem [{\citenamefont {{\v S}mejkal}\ \emph {et~al.}(2020)\citenamefont {{\v
  S}mejkal}, \citenamefont {Gonz{\'a}lez-Hern{\'a}ndez}, \citenamefont
  {Jungwirth},\ and\ \citenamefont {Sinova}}]{Libor20}%
  \BibitemOpen
  \bibfield  {author} {\bibinfo {author} {\bibfnamefont {L.}~\bibnamefont {{\v
  S}mejkal}}, \bibinfo {author} {\bibfnamefont {R.}~\bibnamefont
  {Gonz{\'a}lez-Hern{\'a}ndez}}, \bibinfo {author} {\bibfnamefont
  {T.}~\bibnamefont {Jungwirth}},\ and\ \bibinfo {author} {\bibfnamefont
  {J.}~\bibnamefont {Sinova}},\ }\bibfield  {title} {\bibinfo {title} {{Crystal
  time-reversal symmetry breaking and spontaneous Hall effect in collinear
  antiferromagnets}},\ }\href {https://doi.org/10.1126/sciadv.aaz8809}
  {\bibfield  {journal} {\bibinfo  {journal} {Sci. Adv.}\ }\textbf {\bibinfo
  {volume} {6}},\ \bibinfo {pages} {aaz8809} (\bibinfo {year}
  {2020})}\BibitemShut {NoStop}%
\bibitem [{\citenamefont {Gonz{\'a}lez-Hern{\'a}ndez}\ \emph
  {et~al.}(2021)\citenamefont {Gonz{\'a}lez-Hern{\'a}ndez}, \citenamefont {{\v
  S}mejkal}, \citenamefont {V{\'y}born{\'y}}, \citenamefont {Yahagi},
  \citenamefont {Sinova}, \citenamefont {Jungwirth},\ and\ \citenamefont {{\v
  Z}elezn{\'y}}}]{Libor21}%
  \BibitemOpen
  \bibfield  {author} {\bibinfo {author} {\bibfnamefont {R.}~\bibnamefont
  {Gonz{\'a}lez-Hern{\'a}ndez}}, \bibinfo {author} {\bibfnamefont
  {L.}~\bibnamefont {{\v S}mejkal}}, \bibinfo {author} {\bibfnamefont
  {K.}~\bibnamefont {V{\'y}born{\'y}}}, \bibinfo {author} {\bibfnamefont
  {Y.}~\bibnamefont {Yahagi}}, \bibinfo {author} {\bibfnamefont
  {J.}~\bibnamefont {Sinova}}, \bibinfo {author} {\bibfnamefont
  {T.}~\bibnamefont {Jungwirth}},\ and\ \bibinfo {author} {\bibfnamefont
  {J.}~\bibnamefont {{\v Z}elezn{\'y}}},\ }\bibfield  {title} {\bibinfo {title}
  {{Efficient Electrical Spin Splitter Based on Nonrelativistic Collinear
  Antiferromagnetism}},\ }\href
  {https://doi.org/10.1103/physrevlett.126.127701} {\bibfield  {journal}
  {\bibinfo  {journal} {Phys. Rev. Lett.}\ }\textbf {\bibinfo {volume} {126}},\
  \bibinfo {pages} {127701} (\bibinfo {year} {2021})}\BibitemShut {NoStop}%
\bibitem [{\citenamefont {Feng}\ \emph {et~al.}(2022)\citenamefont {Feng},
  \citenamefont {Zhou}, \citenamefont {{\v S}mejkal}, \citenamefont {Wu},
  \citenamefont {Zhu}, \citenamefont {Guo}, \citenamefont
  {Gonz{\'a}lez-Hern{\'a}ndez}, \citenamefont {Wang}, \citenamefont {Yan},
  \citenamefont {Qin}, \citenamefont {Zhang}, \citenamefont {Wu}, \citenamefont
  {Chen}, \citenamefont {Meng}, \citenamefont {Liu}, \citenamefont {Xia},
  \citenamefont {Sinova}, \citenamefont {Jungwirth},\ and\ \citenamefont
  {Liu}}]{RuO-22}%
  \BibitemOpen
  \bibfield  {author} {\bibinfo {author} {\bibfnamefont {Z.}~\bibnamefont
  {Feng}}, \bibinfo {author} {\bibfnamefont {X.}~\bibnamefont {Zhou}}, \bibinfo
  {author} {\bibfnamefont {L.}~\bibnamefont {{\v S}mejkal}}, \bibinfo {author}
  {\bibfnamefont {L.}~\bibnamefont {Wu}}, \bibinfo {author} {\bibfnamefont
  {Z.}~\bibnamefont {Zhu}}, \bibinfo {author} {\bibfnamefont {H.}~\bibnamefont
  {Guo}}, \bibinfo {author} {\bibfnamefont {R.}~\bibnamefont
  {Gonz{\'a}lez-Hern{\'a}ndez}}, \bibinfo {author} {\bibfnamefont
  {X.}~\bibnamefont {Wang}}, \bibinfo {author} {\bibfnamefont {H.}~\bibnamefont
  {Yan}}, \bibinfo {author} {\bibfnamefont {P.}~\bibnamefont {Qin}}, \bibinfo
  {author} {\bibfnamefont {X.}~\bibnamefont {Zhang}}, \bibinfo {author}
  {\bibfnamefont {H.}~\bibnamefont {Wu}}, \bibinfo {author} {\bibfnamefont
  {H.}~\bibnamefont {Chen}}, \bibinfo {author} {\bibfnamefont {Z.}~\bibnamefont
  {Meng}}, \bibinfo {author} {\bibfnamefont {L.}~\bibnamefont {Liu}}, \bibinfo
  {author} {\bibfnamefont {Z.}~\bibnamefont {Xia}}, \bibinfo {author}
  {\bibfnamefont {J.}~\bibnamefont {Sinova}}, \bibinfo {author} {\bibfnamefont
  {T.}~\bibnamefont {Jungwirth}},\ and\ \bibinfo {author} {\bibfnamefont
  {Z.}~\bibnamefont {Liu}},\ }\bibfield  {title} {\bibinfo {title} {{An
  anomalous Hall effect in altermagnetic ruthenium dioxide}},\ }\href
  {https://doi.org/10.1038/s41928-022-00866-z} {\bibfield  {journal} {\bibinfo
  {journal} {Nat. Electron.}\ }\textbf {\bibinfo {volume} {5}},\ \bibinfo
  {pages} {735} (\bibinfo {year} {2022})}\BibitemShut {NoStop}%
\bibitem [{\citenamefont {Gonzalez~Betancourt}\ \emph
  {et~al.}(2023)\citenamefont {Gonzalez~Betancourt}, \citenamefont {Zub{\'a}{\v
  c}}, \citenamefont {Gonzalez-Hernandez}, \citenamefont {Geishendorf},
  \citenamefont {{\v S}ob{\'a}{\v n}}, \citenamefont {Springholz},
  \citenamefont {Olejn{\'\i}k}, \citenamefont {{\v S}mejkal}, \citenamefont
  {Sinova}, \citenamefont {Jungwirth}, \citenamefont {Goennenwein},
  \citenamefont {Thomas}, \citenamefont {Reichlov{\'a}}, \citenamefont {{\v
  Z}elezn{\'y}},\ and\ \citenamefont {Kriegner}}]{Betancourt23}%
  \BibitemOpen
  \bibfield  {author} {\bibinfo {author} {\bibfnamefont {R.~D.}\ \bibnamefont
  {Gonzalez~Betancourt}}, \bibinfo {author} {\bibfnamefont {J.}~\bibnamefont
  {Zub{\'a}{\v c}}}, \bibinfo {author} {\bibfnamefont {R.}~\bibnamefont
  {Gonzalez-Hernandez}}, \bibinfo {author} {\bibfnamefont {K.}~\bibnamefont
  {Geishendorf}}, \bibinfo {author} {\bibfnamefont {Z.}~\bibnamefont {{\v
  S}ob{\'a}{\v n}}}, \bibinfo {author} {\bibfnamefont {G.}~\bibnamefont
  {Springholz}}, \bibinfo {author} {\bibfnamefont {K.}~\bibnamefont
  {Olejn{\'\i}k}}, \bibinfo {author} {\bibfnamefont {L.}~\bibnamefont {{\v
  S}mejkal}}, \bibinfo {author} {\bibfnamefont {J.}~\bibnamefont {Sinova}},
  \bibinfo {author} {\bibfnamefont {T.}~\bibnamefont {Jungwirth}}, \bibinfo
  {author} {\bibfnamefont {S.~T.~B.}\ \bibnamefont {Goennenwein}}, \bibinfo
  {author} {\bibfnamefont {A.}~\bibnamefont {Thomas}}, \bibinfo {author}
  {\bibfnamefont {H.}~\bibnamefont {Reichlov{\'a}}}, \bibinfo {author}
  {\bibfnamefont {J.}~\bibnamefont {{\v Z}elezn{\'y}}},\ and\ \bibinfo {author}
  {\bibfnamefont {D.}~\bibnamefont {Kriegner}},\ }\bibfield  {title} {\bibinfo
  {title} {{Spontaneous Anomalous Hall Effect Arising from an Unconventional
  Compensated Magnetic Phase in a Semiconductor}},\ }\href
  {https://doi.org/10.1103/physrevlett.130.036702} {\bibfield  {journal}
  {\bibinfo  {journal} {Phys. Rev. Lett.}\ }\textbf {\bibinfo {volume} {130}},\
  \bibinfo {pages} {036702} (\bibinfo {year} {2023})}\BibitemShut {NoStop}%
\bibitem [{\citenamefont {Bai}\ \emph {et~al.}(2022)\citenamefont {Bai},
  \citenamefont {Han}, \citenamefont {Feng}, \citenamefont {Zhou},
  \citenamefont {Su}, \citenamefont {Wang}, \citenamefont {Liao}, \citenamefont
  {Zhu}, \citenamefont {Chen}, \citenamefont {Pan}, \citenamefont {Fan},\ and\
  \citenamefont {Song}}]{Bai22}%
  \BibitemOpen
  \bibfield  {author} {\bibinfo {author} {\bibfnamefont {H.}~\bibnamefont
  {Bai}}, \bibinfo {author} {\bibfnamefont {L.}~\bibnamefont {Han}}, \bibinfo
  {author} {\bibfnamefont {X.~Y.}\ \bibnamefont {Feng}}, \bibinfo {author}
  {\bibfnamefont {Y.~J.}\ \bibnamefont {Zhou}}, \bibinfo {author}
  {\bibfnamefont {R.~X.}\ \bibnamefont {Su}}, \bibinfo {author} {\bibfnamefont
  {Q.}~\bibnamefont {Wang}}, \bibinfo {author} {\bibfnamefont {L.~Y.}\
  \bibnamefont {Liao}}, \bibinfo {author} {\bibfnamefont {W.~X.}\ \bibnamefont
  {Zhu}}, \bibinfo {author} {\bibfnamefont {X.~Z.}\ \bibnamefont {Chen}},
  \bibinfo {author} {\bibfnamefont {F.}~\bibnamefont {Pan}}, \bibinfo {author}
  {\bibfnamefont {X.~L.}\ \bibnamefont {Fan}},\ and\ \bibinfo {author}
  {\bibfnamefont {C.}~\bibnamefont {Song}},\ }\bibfield  {title} {\bibinfo
  {title} {{Observation of Spin Splitting Torque in a Collinear Antiferromagnet
  RuO$_2$}},\ }\href {https://doi.org/10.1103/physrevlett.128.197202}
  {\bibfield  {journal} {\bibinfo  {journal} {Phys. Rev. Lett.}\ }\textbf
  {\bibinfo {volume} {128}},\ \bibinfo {pages} {197202} (\bibinfo {year}
  {2022})}\BibitemShut {NoStop}%
\bibitem [{\citenamefont {Guo}\ \emph {et~al.}(2023{\natexlab{b}})\citenamefont
  {Guo}, \citenamefont {Liu}, \citenamefont {Janson}, \citenamefont {Fulga},
  \citenamefont {{van den Brink}},\ and\ \citenamefont {Facio}}]{Guo23_1}%
  \BibitemOpen
  \bibfield  {author} {\bibinfo {author} {\bibfnamefont {Y.}~\bibnamefont
  {Guo}}, \bibinfo {author} {\bibfnamefont {H.}~\bibnamefont {Liu}}, \bibinfo
  {author} {\bibfnamefont {O.}~\bibnamefont {Janson}}, \bibinfo {author}
  {\bibfnamefont {I.~C.}\ \bibnamefont {Fulga}}, \bibinfo {author}
  {\bibfnamefont {J.}~\bibnamefont {{van den Brink}}},\ and\ \bibinfo {author}
  {\bibfnamefont {J.~I.}\ \bibnamefont {Facio}},\ }\bibfield  {title} {\bibinfo
  {title} {Spin-split collinear antiferromagnets: A large-scale ab-initio
  study},\ }\href
  {https://doi.org/https://doi.org/10.1016/j.mtphys.2023.100991} {\bibfield
  {journal} {\bibinfo  {journal} {Mater. Today Phys.}\ }\textbf {\bibinfo
  {volume} {32}},\ \bibinfo {pages} {100991} (\bibinfo {year}
  {2023}{\natexlab{b}})}\BibitemShut {NoStop}%
\bibitem [{\citenamefont {Reimers}\ \emph {et~al.}(2024)\citenamefont
  {Reimers}, \citenamefont {Odenbreit}, \citenamefont {{\v S}mejkal},
  \citenamefont {Strocov}, \citenamefont {Constantinou}, \citenamefont
  {Hellenes}, \citenamefont {Jaeschke~Ubiergo}, \citenamefont {Campos},
  \citenamefont {Bharadwaj}, \citenamefont {Chakraborty}, \citenamefont
  {Denneulin}, \citenamefont {Shi}, \citenamefont {Dunin-Borkowski},
  \citenamefont {Das}, \citenamefont {Kl{\"a}ui}, \citenamefont {Sinova},\ and\
  \citenamefont {Jourdan}}]{AM-obs}%
  \BibitemOpen
  \bibfield  {author} {\bibinfo {author} {\bibfnamefont {S.}~\bibnamefont
  {Reimers}}, \bibinfo {author} {\bibfnamefont {L.}~\bibnamefont {Odenbreit}},
  \bibinfo {author} {\bibfnamefont {L.}~\bibnamefont {{\v S}mejkal}}, \bibinfo
  {author} {\bibfnamefont {V.~N.}\ \bibnamefont {Strocov}}, \bibinfo {author}
  {\bibfnamefont {P.}~\bibnamefont {Constantinou}}, \bibinfo {author}
  {\bibfnamefont {A.~B.}\ \bibnamefont {Hellenes}}, \bibinfo {author}
  {\bibfnamefont {R.}~\bibnamefont {Jaeschke~Ubiergo}}, \bibinfo {author}
  {\bibfnamefont {W.~H.}\ \bibnamefont {Campos}}, \bibinfo {author}
  {\bibfnamefont {V.~K.}\ \bibnamefont {Bharadwaj}}, \bibinfo {author}
  {\bibfnamefont {A.}~\bibnamefont {Chakraborty}}, \bibinfo {author}
  {\bibfnamefont {T.}~\bibnamefont {Denneulin}}, \bibinfo {author}
  {\bibfnamefont {W.}~\bibnamefont {Shi}}, \bibinfo {author} {\bibfnamefont
  {R.~E.}\ \bibnamefont {Dunin-Borkowski}}, \bibinfo {author} {\bibfnamefont
  {S.}~\bibnamefont {Das}}, \bibinfo {author} {\bibfnamefont {M.}~\bibnamefont
  {Kl{\"a}ui}}, \bibinfo {author} {\bibfnamefont {J.}~\bibnamefont {Sinova}},\
  and\ \bibinfo {author} {\bibfnamefont {M.}~\bibnamefont {Jourdan}},\
  }\bibfield  {title} {\bibinfo {title} {{Direct observation of altermagnetic
  band splitting in CrSb thin films}},\ }\href
  {https://doi.org/10.1038/s41467-024-46476-5} {\bibfield  {journal} {\bibinfo
  {journal} {Nat. Commun.}\ }\textbf {\bibinfo {volume} {15}},\ \bibinfo
  {pages} {2116} (\bibinfo {year} {2024})}\BibitemShut {NoStop}%
\bibitem [{\citenamefont {Sato}\ \emph {et~al.}(2024)\citenamefont {Sato},
  \citenamefont {Haddad}, \citenamefont {Fulga}, \citenamefont {Assaad},\ and\
  \citenamefont {van~den Brink}}]{SatoT24_1}%
  \BibitemOpen
  \bibfield  {author} {\bibinfo {author} {\bibfnamefont {T.}~\bibnamefont
  {Sato}}, \bibinfo {author} {\bibfnamefont {S.}~\bibnamefont {Haddad}},
  \bibinfo {author} {\bibfnamefont {I.~C.}\ \bibnamefont {Fulga}}, \bibinfo
  {author} {\bibfnamefont {F.~F.}\ \bibnamefont {Assaad}},\ and\ \bibinfo
  {author} {\bibfnamefont {J.}~\bibnamefont {van~den Brink}},\ }\bibfield
  {title} {\bibinfo {title} {{Altermagnetic Anomalous Hall Effect Emerging from
  Electronic Correlations}},\ }\href
  {https://doi.org/10.1103/physrevlett.133.086503} {\bibfield  {journal}
  {\bibinfo  {journal} {Phys. Rev. Lett.}\ }\textbf {\bibinfo {volume} {133}},\
  \bibinfo {pages} {086503} (\bibinfo {year} {2024})}\BibitemShut {NoStop}%
\bibitem [{\citenamefont {Li}\ \emph {et~al.}(2025)\citenamefont {Li},
  \citenamefont {Hu}, \citenamefont {Li}, \citenamefont {Wang}, \citenamefont
  {Chen}, \citenamefont {Thiagarajan}, \citenamefont {Leandersson},
  \citenamefont {Polley}, \citenamefont {Kim}, \citenamefont {Liu},
  \citenamefont {Fulga}, \citenamefont {Vergniory}, \citenamefont {Janson},
  \citenamefont {Tjernberg},\ and\ \citenamefont {van~den Brink}}]{Li_24_1}%
  \BibitemOpen
  \bibfield  {author} {\bibinfo {author} {\bibfnamefont {C.}~\bibnamefont
  {Li}}, \bibinfo {author} {\bibfnamefont {M.}~\bibnamefont {Hu}}, \bibinfo
  {author} {\bibfnamefont {Z.}~\bibnamefont {Li}}, \bibinfo {author}
  {\bibfnamefont {Y.}~\bibnamefont {Wang}}, \bibinfo {author} {\bibfnamefont
  {W.}~\bibnamefont {Chen}}, \bibinfo {author} {\bibfnamefont {B.}~\bibnamefont
  {Thiagarajan}}, \bibinfo {author} {\bibfnamefont {M.}~\bibnamefont
  {Leandersson}}, \bibinfo {author} {\bibfnamefont {C.}~\bibnamefont {Polley}},
  \bibinfo {author} {\bibfnamefont {T.}~\bibnamefont {Kim}}, \bibinfo {author}
  {\bibfnamefont {H.}~\bibnamefont {Liu}}, \bibinfo {author} {\bibfnamefont
  {C.}~\bibnamefont {Fulga}}, \bibinfo {author} {\bibfnamefont {M.~G.}\
  \bibnamefont {Vergniory}}, \bibinfo {author} {\bibfnamefont {O.}~\bibnamefont
  {Janson}}, \bibinfo {author} {\bibfnamefont {O.}~\bibnamefont {Tjernberg}},\
  and\ \bibinfo {author} {\bibfnamefont {J.}~\bibnamefont {van~den Brink}},\
  }\bibfield  {title} {\bibinfo {title} {{Topological Weyl altermagnetism in
  CrSb}},\ }\href {https://doi.org/10.1038/s42005-025-02232-9} {\bibfield
  {journal} {\bibinfo  {journal} {Commun. Phys.}\ }\textbf {\bibinfo {volume}
  {8}},\ \bibinfo {pages} {311} (\bibinfo {year} {2025})}\BibitemShut {NoStop}%
\bibitem [{\citenamefont {Hu}\ \emph {et~al.}(2024)\citenamefont {Hu},
  \citenamefont {Janson}, \citenamefont {Felser}, \citenamefont {McClarty},
  \citenamefont {van~den Brink},\ and\ \citenamefont {Vergniory}}]{Hu_24_1}%
  \BibitemOpen
  \bibfield  {author} {\bibinfo {author} {\bibfnamefont {M.}~\bibnamefont
  {Hu}}, \bibinfo {author} {\bibfnamefont {O.}~\bibnamefont {Janson}}, \bibinfo
  {author} {\bibfnamefont {C.}~\bibnamefont {Felser}}, \bibinfo {author}
  {\bibfnamefont {P.}~\bibnamefont {McClarty}}, \bibinfo {author}
  {\bibfnamefont {J.}~\bibnamefont {van~den Brink}},\ and\ \bibinfo {author}
  {\bibfnamefont {M.~G.}\ \bibnamefont {Vergniory}},\ }\bibfield  {title}
  {\bibinfo {title} {{Spin Hall and Edelstein Effects in Novel Chiral
  Noncollinear Altermagnets}},\ }\href {https://arxiv.org/abs/2410.17993}
  {\bibfield  {journal} {\bibinfo  {journal} {arXiv:2410.1799}\ } (\bibinfo
  {year} {2024})}\BibitemShut {NoStop}%
\bibitem [{\citenamefont {Trama}\ \emph {et~al.}(2024)\citenamefont {Trama},
  \citenamefont {Gaiardoni}, \citenamefont {Guarcello}, \citenamefont {Facio},
  \citenamefont {Maiellaro}, \citenamefont {Romeo}, \citenamefont {Citro},\
  and\ \citenamefont {van~den Brink}}]{Trama_24_1}%
  \BibitemOpen
  \bibfield  {author} {\bibinfo {author} {\bibfnamefont {M.}~\bibnamefont
  {Trama}}, \bibinfo {author} {\bibfnamefont {I.}~\bibnamefont {Gaiardoni}},
  \bibinfo {author} {\bibfnamefont {C.}~\bibnamefont {Guarcello}}, \bibinfo
  {author} {\bibfnamefont {J.~I.}\ \bibnamefont {Facio}}, \bibinfo {author}
  {\bibfnamefont {A.}~\bibnamefont {Maiellaro}}, \bibinfo {author}
  {\bibfnamefont {F.}~\bibnamefont {Romeo}}, \bibinfo {author} {\bibfnamefont
  {R.}~\bibnamefont {Citro}},\ and\ \bibinfo {author} {\bibfnamefont
  {J.}~\bibnamefont {van~den Brink}},\ }\bibfield  {title} {\bibinfo {title}
  {{Non-linear anomalous Edelstein response at altermagnetic interfaces}},\
  }\href {https://arxiv.org/abs/2410.18036} {\bibfield  {journal} {\bibinfo
  {journal} {arXiv:2410.18036}\ } (\bibinfo {year} {2024})}\BibitemShut
  {NoStop}%
\bibitem [{\citenamefont {Yershov}\ \emph {et~al.}(2024)\citenamefont
  {Yershov}, \citenamefont {Kravchuk}, \citenamefont {Daghofer},\ and\
  \citenamefont {van~den Brink}}]{Yershov24_1}%
  \BibitemOpen
  \bibfield  {author} {\bibinfo {author} {\bibfnamefont {K.~V.}\ \bibnamefont
  {Yershov}}, \bibinfo {author} {\bibfnamefont {V.~P.}\ \bibnamefont
  {Kravchuk}}, \bibinfo {author} {\bibfnamefont {M.}~\bibnamefont {Daghofer}},\
  and\ \bibinfo {author} {\bibfnamefont {J.}~\bibnamefont {van~den Brink}},\
  }\bibfield  {title} {\bibinfo {title} {Fluctuation-induced piezomagnetism in
  local moment altermagnets},\ }\href
  {https://doi.org/10.1103/PhysRevB.110.144421} {\bibfield  {journal} {\bibinfo
   {journal} {Phys. Rev. B}\ }\textbf {\bibinfo {volume} {110}},\ \bibinfo
  {pages} {144421} (\bibinfo {year} {2024})}\BibitemShut {NoStop}%
\bibitem [{\citenamefont {Kravchuk}\ \emph {et~al.}(2025)\citenamefont
  {Kravchuk}, \citenamefont {Yershov}, \citenamefont {Facio}, \citenamefont
  {Guo}, \citenamefont {Janson}, \citenamefont {Gomonay}, \citenamefont
  {Sinova},\ and\ \citenamefont {van~den Brink}}]{Kravchuk_25_1}%
  \BibitemOpen
  \bibfield  {author} {\bibinfo {author} {\bibfnamefont {V.~P.}\ \bibnamefont
  {Kravchuk}}, \bibinfo {author} {\bibfnamefont {K.~V.}\ \bibnamefont
  {Yershov}}, \bibinfo {author} {\bibfnamefont {J.~I.}\ \bibnamefont {Facio}},
  \bibinfo {author} {\bibfnamefont {Y.}~\bibnamefont {Guo}}, \bibinfo {author}
  {\bibfnamefont {O.}~\bibnamefont {Janson}}, \bibinfo {author} {\bibfnamefont
  {O.}~\bibnamefont {Gomonay}}, \bibinfo {author} {\bibfnamefont
  {J.}~\bibnamefont {Sinova}},\ and\ \bibinfo {author} {\bibfnamefont
  {J.}~\bibnamefont {van~den Brink}},\ }\bibfield  {title} {\bibinfo {title}
  {Chiral magnetic excitations and domain textures of g-wave altermagnets},\
  }\href {https://arxiv.org/abs/2504.05241} {\bibfield  {journal} {\bibinfo
  {journal} {arXiv:2504.05241}\ } (\bibinfo {year} {2025})}\BibitemShut
  {NoStop}%
\bibitem [{Note1()}]{Note1}%
  \BibitemOpen
  \bibinfo {note} {Generalization to the case with finite spin-orbit coupling
  is straightforward.}\BibitemShut {Stop}%
\bibitem [{\citenamefont {King-Smith}\ and\ \citenamefont
  {Vanderbilt}(1993)}]{King-Smith1993}%
  \BibitemOpen
  \bibfield  {author} {\bibinfo {author} {\bibfnamefont {R.~D.}\ \bibnamefont
  {King-Smith}}\ and\ \bibinfo {author} {\bibfnamefont {D.}~\bibnamefont
  {Vanderbilt}},\ }\bibfield  {title} {\bibinfo {title} {Theory of polarization
  of crystalline solids},\ }\href {https://doi.org/10.1103/PhysRevB.47.1651}
  {\bibfield  {journal} {\bibinfo  {journal} {Phys. Rev. B}\ }\textbf {\bibinfo
  {volume} {47}},\ \bibinfo {pages} {1651} (\bibinfo {year}
  {1993})}\BibitemShut {NoStop}%
\bibitem [{\citenamefont {Resta}(1994)}]{Resta1994}%
  \BibitemOpen
  \bibfield  {author} {\bibinfo {author} {\bibfnamefont {R.}~\bibnamefont
  {Resta}},\ }\bibfield  {title} {\bibinfo {title} {Macroscopic polarization in
  crystalline dielectrics: the geometric phase approach},\ }\href
  {https://doi.org/10.1103/RevModPhys.66.899} {\bibfield  {journal} {\bibinfo
  {journal} {Rev. Mod. Phys.}\ }\textbf {\bibinfo {volume} {66}},\ \bibinfo
  {pages} {899} (\bibinfo {year} {1994})}\BibitemShut {NoStop}%
\bibitem [{\citenamefont {Resta}(1998)}]{Resta1998}%
  \BibitemOpen
  \bibfield  {author} {\bibinfo {author} {\bibfnamefont {R.}~\bibnamefont
  {Resta}},\ }\bibfield  {title} {\bibinfo {title} {Quantum-mechanical position
  operator in extended systems},\ }\href
  {https://doi.org/10.1103/PhysRevLett.80.1800} {\bibfield  {journal} {\bibinfo
   {journal} {Phys. Rev. Lett.}\ }\textbf {\bibinfo {volume} {80}},\ \bibinfo
  {pages} {1800} (\bibinfo {year} {1998})}\BibitemShut {NoStop}%
\bibitem [{Note2()}]{Note2}%
  \BibitemOpen
  \bibinfo {note} {Ref.~\protect \rev@citealp {Saez2025} coins the term
  ferro-spintronic order for the dipole moment of the spin density. As
  spintronics is an already established field of technology, we prefer not to
  conflate our nomenclature with that terminology.}\BibitemShut {Stop}%
\bibitem [{\citenamefont {Shi}\ \emph {et~al.}(2006)\citenamefont {Shi},
  \citenamefont {Zhang}, \citenamefont {Xiao},\ and\ \citenamefont
  {Niu}}]{Shi2006}%
  \BibitemOpen
  \bibfield  {author} {\bibinfo {author} {\bibfnamefont {J.}~\bibnamefont
  {Shi}}, \bibinfo {author} {\bibfnamefont {P.}~\bibnamefont {Zhang}}, \bibinfo
  {author} {\bibfnamefont {D.}~\bibnamefont {Xiao}},\ and\ \bibinfo {author}
  {\bibfnamefont {Q.}~\bibnamefont {Niu}},\ }\bibfield  {title} {\bibinfo
  {title} {Proper definition of spin current in spin-orbit coupled systems},\
  }\href {https://doi.org/10.1103/PhysRevLett.96.076604} {\bibfield  {journal}
  {\bibinfo  {journal} {Phys. Rev. Lett.}\ }\textbf {\bibinfo {volume} {96}},\
  \bibinfo {pages} {076604} (\bibinfo {year} {2006})}\BibitemShut {NoStop}%
\bibitem [{\citenamefont {Bercx}\ \emph {et~al.}(2017)\citenamefont {Bercx},
  \citenamefont {Goth}, \citenamefont {Hofmann},\ and\ \citenamefont
  {Assaad}}]{ALF_v1}%
  \BibitemOpen
  \bibfield  {author} {\bibinfo {author} {\bibfnamefont {M.}~\bibnamefont
  {Bercx}}, \bibinfo {author} {\bibfnamefont {F.}~\bibnamefont {Goth}},
  \bibinfo {author} {\bibfnamefont {J.~S.}\ \bibnamefont {Hofmann}},\ and\
  \bibinfo {author} {\bibfnamefont {F.}~\bibnamefont {Assaad}},\ }\bibfield
  {title} {\bibinfo {title} {{The ALF (Algorithms for Lattice Fermions) project
  release 1.0. Documentation for the auxiliary field quantum Monte Carlo
  code}},\ }\href {https://doi.org/10.21468/scipostphys.3.2.013} {\bibfield
  {journal} {\bibinfo  {journal} {SciPost Phys.}\ }\textbf {\bibinfo {volume}
  {3}},\ \bibinfo {pages} {013} (\bibinfo {year} {2017})}\BibitemShut {NoStop}%
\bibitem [{\citenamefont {Assaad}\ \emph {et~al.}(2022)\citenamefont {Assaad},
  \citenamefont {Bercx}, \citenamefont {Goth}, \citenamefont {G{\"o}tz},
  \citenamefont {Hofmann}, \citenamefont {Huffman}, \citenamefont {Liu},
  \citenamefont {Parisen~Toldin}, \citenamefont {Portela},\ and\ \citenamefont
  {Schwab}}]{ALF_v2}%
  \BibitemOpen
  \bibfield  {author} {\bibinfo {author} {\bibfnamefont {F.}~\bibnamefont
  {Assaad}}, \bibinfo {author} {\bibfnamefont {M.}~\bibnamefont {Bercx}},
  \bibinfo {author} {\bibfnamefont {F.}~\bibnamefont {Goth}}, \bibinfo {author}
  {\bibfnamefont {A.}~\bibnamefont {G{\"o}tz}}, \bibinfo {author}
  {\bibfnamefont {J.}~\bibnamefont {Hofmann}}, \bibinfo {author} {\bibfnamefont
  {E.}~\bibnamefont {Huffman}}, \bibinfo {author} {\bibfnamefont
  {Z.}~\bibnamefont {Liu}}, \bibinfo {author} {\bibfnamefont {F.}~\bibnamefont
  {Parisen~Toldin}}, \bibinfo {author} {\bibfnamefont {J.}~\bibnamefont
  {Portela}},\ and\ \bibinfo {author} {\bibfnamefont {J.}~\bibnamefont
  {Schwab}},\ }\bibfield  {title} {\bibinfo {title} {{The ALF (Algorithms for
  Lattice Fermions) project release 2.0. Documentation for the auxiliary-field
  quantum Monte Carlo code}},\ }\href
  {https://doi.org/10.21468/scipostphyscodeb.1} {\bibfield  {journal} {\bibinfo
   {journal} {SciPost Phys. Codebases}\ ,\ \bibinfo {pages} {1}} (\bibinfo
  {year} {2022})}\BibitemShut {NoStop}%
\bibitem [{Note3()}]{Note3}%
  \BibitemOpen
  \bibinfo {note} {{Including $\eta $ term breaks chiral symmetry and
  introduces a negative sign problem; in the $\eta $ range explored here it
  remains mild (e.g., at $\eta =0.008$ we find quantum Monte Carlo average
  sign, $\delimiter "426830A \protect \text {sign} \delimiter "526930B \approx
  0.606$)}}\BibitemShut {NoStop}%
\bibitem [{\citenamefont {Di~Sante}\ \emph {et~al.}(2013)\citenamefont
  {Di~Sante}, \citenamefont {Stroppa}, \citenamefont {Jain},\ and\
  \citenamefont {Picozzi}}]{Mn-MOF_numerical}%
  \BibitemOpen
  \bibfield  {author} {\bibinfo {author} {\bibfnamefont {D.}~\bibnamefont
  {Di~Sante}}, \bibinfo {author} {\bibfnamefont {A.}~\bibnamefont {Stroppa}},
  \bibinfo {author} {\bibfnamefont {P.}~\bibnamefont {Jain}},\ and\ \bibinfo
  {author} {\bibfnamefont {S.}~\bibnamefont {Picozzi}},\ }\bibfield  {title}
  {\bibinfo {title} {Tuning the ferroelectric polarization in a multiferroic
  metal--organic framework},\ }\href {https://doi.org/10.1021/ja408283a}
  {\bibfield  {journal} {\bibinfo  {journal} {J. Am. Chem. Soc.}\ }\textbf
  {\bibinfo {volume} {135}},\ \bibinfo {pages} {18126} (\bibinfo {year}
  {2013})}\BibitemShut {NoStop}%
\bibitem [{\citenamefont {Knight}\ \emph {et~al.}(2020)\citenamefont {Knight},
  \citenamefont {Khalyavin}, \citenamefont {Manuel}, \citenamefont {Bull},\
  and\ \citenamefont {McIntyre}}]{KNIGHT2020155935_kmnf3}%
  \BibitemOpen
  \bibfield  {author} {\bibinfo {author} {\bibfnamefont {K.~S.}\ \bibnamefont
  {Knight}}, \bibinfo {author} {\bibfnamefont {D.~D.}\ \bibnamefont
  {Khalyavin}}, \bibinfo {author} {\bibfnamefont {P.}~\bibnamefont {Manuel}},
  \bibinfo {author} {\bibfnamefont {C.~L.}\ \bibnamefont {Bull}},\ and\
  \bibinfo {author} {\bibfnamefont {P.}~\bibnamefont {McIntyre}},\ }\bibfield
  {title} {\bibinfo {title} {Nuclear and magnetic structures of kmnf3
  perovskite in the temperature interval 10 k--105 k},\ }\href
  {https://doi.org/https://doi.org/10.1016/j.jallcom.2020.155935} {\bibfield
  {journal} {\bibinfo  {journal} {Journal of Alloys and Compounds}\ }\textbf
  {\bibinfo {volume} {842}},\ \bibinfo {pages} {155935} (\bibinfo {year}
  {2020})}\BibitemShut {NoStop}%
\bibitem [{\citenamefont {Blankenbecler}\ \emph {et~al.}(1981)\citenamefont
  {Blankenbecler}, \citenamefont {Scalapino},\ and\ \citenamefont
  {Sugar}}]{Blankenbecler81}%
  \BibitemOpen
  \bibfield  {author} {\bibinfo {author} {\bibfnamefont {R.}~\bibnamefont
  {Blankenbecler}}, \bibinfo {author} {\bibfnamefont {D.~J.}\ \bibnamefont
  {Scalapino}},\ and\ \bibinfo {author} {\bibfnamefont {R.~L.}\ \bibnamefont
  {Sugar}},\ }\bibfield  {title} {\bibinfo {title} {{Monte Carlo calculations
  of coupled boson-fermion systems. I}},\ }\href
  {https://doi.org/10.1103/physrevd.24.2278} {\bibfield  {journal} {\bibinfo
  {journal} {Phys. Rev. D}\ }\textbf {\bibinfo {volume} {24}},\ \bibinfo
  {pages} {2278} (\bibinfo {year} {1981})}\BibitemShut {NoStop}%
\bibitem [{\citenamefont {White}\ \emph {et~al.}(1989)\citenamefont {White},
  \citenamefont {Scalapino}, \citenamefont {Sugar}, \citenamefont {Loh},
  \citenamefont {Gubernatis},\ and\ \citenamefont {Scalettar}}]{White89}%
  \BibitemOpen
  \bibfield  {author} {\bibinfo {author} {\bibfnamefont {S.~R.}\ \bibnamefont
  {White}}, \bibinfo {author} {\bibfnamefont {D.~J.}\ \bibnamefont
  {Scalapino}}, \bibinfo {author} {\bibfnamefont {R.~L.}\ \bibnamefont
  {Sugar}}, \bibinfo {author} {\bibfnamefont {E.~Y.}\ \bibnamefont {Loh}},
  \bibinfo {author} {\bibfnamefont {J.~E.}\ \bibnamefont {Gubernatis}},\ and\
  \bibinfo {author} {\bibfnamefont {R.~T.}\ \bibnamefont {Scalettar}},\
  }\bibfield  {title} {\bibinfo {title} {{Numerical study of the
  two-dimensional Hubbard model}},\ }\href
  {https://doi.org/10.1103/physrevb.40.506} {\bibfield  {journal} {\bibinfo
  {journal} {Phys. Rev. B}\ }\textbf {\bibinfo {volume} {40}},\ \bibinfo
  {pages} {506} (\bibinfo {year} {1989})}\BibitemShut {NoStop}%
\bibitem [{\citenamefont {Assaad}\ and\ \citenamefont
  {Evertz}(2008)}]{Assaad08_rev}%
  \BibitemOpen
  \bibfield  {author} {\bibinfo {author} {\bibfnamefont {F.}~\bibnamefont
  {Assaad}}\ and\ \bibinfo {author} {\bibfnamefont {H.}~\bibnamefont
  {Evertz}},\ }\bibinfo {title} {{World-line and Determinantal Quantum Monte
  Carlo Methods for Spins, Phonons and Electrons}},\ in\ \href
  {https://doi.org/10.1007/978-3-540-74686-7} {\emph {\bibinfo {booktitle}
  {Computational Many-Particle Physics}}}\ (\bibinfo  {publisher} {Springer
  Berlin Heidelberg},\ \bibinfo {year} {2008})\ pp.\ \bibinfo {pages}
  {277--356}\BibitemShut {NoStop}%
\bibitem [{\citenamefont {Kresse}\ and\ \citenamefont
  {Furthm\"uller}(1996)}]{GKresse_PRB1996_JFurthmuller}%
  \BibitemOpen
  \bibfield  {author} {\bibinfo {author} {\bibfnamefont {G.}~\bibnamefont
  {Kresse}}\ and\ \bibinfo {author} {\bibfnamefont {J.}~\bibnamefont
  {Furthm\"uller}},\ }\bibfield  {title} {\bibinfo {title} {Efficient iterative
  schemes for ab initio total-energy calculations using a plane-wave basis
  set},\ }\href {https://doi.org/10.1103/PhysRevB.54.11169} {\bibfield
  {journal} {\bibinfo  {journal} {Phys. Rev. B}\ }\textbf {\bibinfo {volume}
  {54}},\ \bibinfo {pages} {11169} (\bibinfo {year} {1996})}\BibitemShut
  {NoStop}%
\bibitem [{\citenamefont {Bl\"ochl}(1994)}]{PEBlochl_PRB1994}%
  \BibitemOpen
  \bibfield  {author} {\bibinfo {author} {\bibfnamefont {P.~E.}\ \bibnamefont
  {Bl\"ochl}},\ }\bibfield  {title} {\bibinfo {title} {Projector augmented-wave
  method},\ }\href {https://doi.org/10.1103/PhysRevB.50.17953} {\bibfield
  {journal} {\bibinfo  {journal} {Phys. Rev. B}\ }\textbf {\bibinfo {volume}
  {50}},\ \bibinfo {pages} {17953} (\bibinfo {year} {1994})}\BibitemShut
  {NoStop}%
\bibitem [{\citenamefont {Saez}\ \emph {et~al.}(2025)\citenamefont {Saez},
  \citenamefont {Vergara}, \citenamefont {Castro}, \citenamefont {Allende},\
  and\ \citenamefont {Nunez}}]{Saez2025}%
  \BibitemOpen
  \bibfield  {author} {\bibinfo {author} {\bibfnamefont {G.}~\bibnamefont
  {Saez}}, \bibinfo {author} {\bibfnamefont {P.}~\bibnamefont {Vergara}},
  \bibinfo {author} {\bibfnamefont {M.}~\bibnamefont {Castro}}, \bibinfo
  {author} {\bibfnamefont {S.}~\bibnamefont {Allende}},\ and\ \bibinfo {author}
  {\bibfnamefont {A.~S.}\ \bibnamefont {Nunez}},\ }\bibfield  {title} {\bibinfo
  {title} {Ferrospintronic order in noncentrosymmetric antiferromagnets: An
  avenue toward spintronic-based computing, data storage, and energy
  harvesting},\ }\href {https://doi.org/https://doi.org/10.1002/pssr.202400292}
  {\bibfield  {journal} {\bibinfo  {journal} {physica status solidi (RRL) --
  Rapid Research Letters}\ }\textbf {\bibinfo {volume} {19}},\ \bibinfo {pages}
  {2400292} (\bibinfo {year} {2025})}\BibitemShut {NoStop}%
\bibitem [{\citenamefont {Gurarie}(2011)}]{Gurarie2011}%
  \BibitemOpen
  \bibfield  {author} {\bibinfo {author} {\bibfnamefont {V.}~\bibnamefont
  {Gurarie}},\ }\bibfield  {title} {\bibinfo {title} {Single-particle green's
  functions and interacting topological insulators},\ }\href
  {https://doi.org/10.1103/PhysRevB.83.085426} {\bibfield  {journal} {\bibinfo
  {journal} {Phys. Rev. B}\ }\textbf {\bibinfo {volume} {83}},\ \bibinfo
  {pages} {085426} (\bibinfo {year} {2011})}\BibitemShut {NoStop}%
\bibitem [{\citenamefont {Kimura}\ \emph {et~al.}(2019)\citenamefont {Kimura},
  \citenamefont {Yoshida},\ and\ \citenamefont {Kawakami}}]{Kimura2019}%
  \BibitemOpen
  \bibfield  {author} {\bibinfo {author} {\bibfnamefont {K.}~\bibnamefont
  {Kimura}}, \bibinfo {author} {\bibfnamefont {T.}~\bibnamefont {Yoshida}},\
  and\ \bibinfo {author} {\bibfnamefont {N.}~\bibnamefont {Kawakami}},\
  }\bibfield  {title} {\bibinfo {title} {Chiral-symmetry protected exceptional
  torus in correlated nodal-line semimetals},\ }\href
  {https://doi.org/10.1103/PhysRevB.100.115124} {\bibfield  {journal} {\bibinfo
   {journal} {Phys. Rev. B}\ }\textbf {\bibinfo {volume} {100}},\ \bibinfo
  {pages} {115124} (\bibinfo {year} {2019})}\BibitemShut {NoStop}%
\bibitem [{\citenamefont {Binder}(1981)}]{Binder1981}%
  \BibitemOpen
  \bibfield  {author} {\bibinfo {author} {\bibfnamefont {K.}~\bibnamefont
  {Binder}},\ }\bibfield  {title} {\bibinfo {title} {{Finite size scaling
  analysis of ising model block distribution functions}},\ }\href
  {https://doi.org/10.1007/bf01293604} {\bibfield  {journal} {\bibinfo
  {journal} {Z. Phys. B Con. Mat.}\ }\textbf {\bibinfo {volume} {43}},\
  \bibinfo {pages} {119} (\bibinfo {year} {1981})}\BibitemShut {NoStop}%
\bibitem [{\citenamefont {Pujari}\ \emph {et~al.}(2016)\citenamefont {Pujari},
  \citenamefont {Lang}, \citenamefont {Murthy},\ and\ \citenamefont
  {Kaul}}]{Pujari16}%
  \BibitemOpen
  \bibfield  {author} {\bibinfo {author} {\bibfnamefont {S.}~\bibnamefont
  {Pujari}}, \bibinfo {author} {\bibfnamefont {T.~C.}\ \bibnamefont {Lang}},
  \bibinfo {author} {\bibfnamefont {G.}~\bibnamefont {Murthy}},\ and\ \bibinfo
  {author} {\bibfnamefont {R.~K.}\ \bibnamefont {Kaul}},\ }\bibfield  {title}
  {\bibinfo {title} {{Interaction-Induced Dirac Fermions from Quadratic Band
  Touching in Bilayer Graphene}},\ }\href
  {https://doi.org/10.1103/physrevlett.117.086404} {\bibfield  {journal}
  {\bibinfo  {journal} {Phys. Rev. Lett.}\ }\textbf {\bibinfo {volume} {117}},\
  \bibinfo {pages} {086404} (\bibinfo {year} {2016})}\BibitemShut {NoStop}%
\bibitem [{\citenamefont {Assaad}\ and\ \citenamefont
  {Herbut}(2013)}]{Assaad13}%
  \BibitemOpen
  \bibfield  {author} {\bibinfo {author} {\bibfnamefont {F.~F.}\ \bibnamefont
  {Assaad}}\ and\ \bibinfo {author} {\bibfnamefont {I.~F.}\ \bibnamefont
  {Herbut}},\ }\bibfield  {title} {\bibinfo {title} {Pinning the order: The
  nature of quantum criticality in the hubbard model on honeycomb lattice},\
  }\href {https://doi.org/10.1103/PhysRevX.3.031010} {\bibfield  {journal}
  {\bibinfo  {journal} {Phys. Rev. X}\ }\textbf {\bibinfo {volume} {3}},\
  \bibinfo {pages} {031010} (\bibinfo {year} {2013})}\BibitemShut {NoStop}%
\bibitem [{\citenamefont {Loh\"ofer}\ \emph {et~al.}(2015)\citenamefont
  {Loh\"ofer}, \citenamefont {Coletta}, \citenamefont {Joshi}, \citenamefont
  {Assaad}, \citenamefont {Vojta}, \citenamefont {Wessel},\ and\ \citenamefont
  {Mila}}]{Lohofer15}%
  \BibitemOpen
  \bibfield  {author} {\bibinfo {author} {\bibfnamefont {M.}~\bibnamefont
  {Loh\"ofer}}, \bibinfo {author} {\bibfnamefont {T.}~\bibnamefont {Coletta}},
  \bibinfo {author} {\bibfnamefont {D.~G.}\ \bibnamefont {Joshi}}, \bibinfo
  {author} {\bibfnamefont {F.~F.}\ \bibnamefont {Assaad}}, \bibinfo {author}
  {\bibfnamefont {M.}~\bibnamefont {Vojta}}, \bibinfo {author} {\bibfnamefont
  {S.}~\bibnamefont {Wessel}},\ and\ \bibinfo {author} {\bibfnamefont
  {F.}~\bibnamefont {Mila}},\ }\bibfield  {title} {\bibinfo {title} {Dynamical
  structure factors and excitation modes of the bilayer heisenberg model},\
  }\href {https://doi.org/10.1103/PhysRevB.92.245137} {\bibfield  {journal}
  {\bibinfo  {journal} {Phys. Rev. B}\ }\textbf {\bibinfo {volume} {92}},\
  \bibinfo {pages} {245137} (\bibinfo {year} {2015})}\BibitemShut {NoStop}%
\bibitem [{\citenamefont {Xiao}\ \emph {et~al.}(2024)\citenamefont {Xiao},
  \citenamefont {Zhao}, \citenamefont {Li}, \citenamefont {Shindou},\ and\
  \citenamefont {Song}}]{SSG1}%
  \BibitemOpen
  \bibfield  {author} {\bibinfo {author} {\bibfnamefont {Z.}~\bibnamefont
  {Xiao}}, \bibinfo {author} {\bibfnamefont {J.}~\bibnamefont {Zhao}}, \bibinfo
  {author} {\bibfnamefont {Y.}~\bibnamefont {Li}}, \bibinfo {author}
  {\bibfnamefont {R.}~\bibnamefont {Shindou}},\ and\ \bibinfo {author}
  {\bibfnamefont {Z.-D.}\ \bibnamefont {Song}},\ }\bibfield  {title} {\bibinfo
  {title} {Spin space groups: Full classification and applications},\ }\href
  {https://doi.org/10.1103/PhysRevX.14.031037} {\bibfield  {journal} {\bibinfo
  {journal} {Phys. Rev. X}\ }\textbf {\bibinfo {volume} {14}},\ \bibinfo
  {pages} {031037} (\bibinfo {year} {2024})}\BibitemShut {NoStop}%
\bibitem [{\citenamefont {Chen}\ \emph {et~al.}(2024)\citenamefont {Chen},
  \citenamefont {Ren}, \citenamefont {Zhu}, \citenamefont {Yu}, \citenamefont
  {Zhang}, \citenamefont {Liu}, \citenamefont {Li}, \citenamefont {Liu},
  \citenamefont {Li},\ and\ \citenamefont {Liu}}]{SSG2}%
  \BibitemOpen
  \bibfield  {author} {\bibinfo {author} {\bibfnamefont {X.}~\bibnamefont
  {Chen}}, \bibinfo {author} {\bibfnamefont {J.}~\bibnamefont {Ren}}, \bibinfo
  {author} {\bibfnamefont {Y.}~\bibnamefont {Zhu}}, \bibinfo {author}
  {\bibfnamefont {Y.}~\bibnamefont {Yu}}, \bibinfo {author} {\bibfnamefont
  {A.}~\bibnamefont {Zhang}}, \bibinfo {author} {\bibfnamefont
  {P.}~\bibnamefont {Liu}}, \bibinfo {author} {\bibfnamefont {J.}~\bibnamefont
  {Li}}, \bibinfo {author} {\bibfnamefont {Y.}~\bibnamefont {Liu}}, \bibinfo
  {author} {\bibfnamefont {C.}~\bibnamefont {Li}},\ and\ \bibinfo {author}
  {\bibfnamefont {Q.}~\bibnamefont {Liu}},\ }\bibfield  {title} {\bibinfo
  {title} {Enumeration and representation theory of spin space groups},\ }\href
  {https://doi.org/10.1103/PhysRevX.14.031038} {\bibfield  {journal} {\bibinfo
  {journal} {Phys. Rev. X}\ }\textbf {\bibinfo {volume} {14}},\ \bibinfo
  {pages} {031038} (\bibinfo {year} {2024})}\BibitemShut {NoStop}%
\bibitem [{\citenamefont {Jiang}\ \emph {et~al.}(2024)\citenamefont {Jiang},
  \citenamefont {Song}, \citenamefont {Zhu}, \citenamefont {Fang},
  \citenamefont {Weng}, \citenamefont {Liu}, \citenamefont {Yang},\ and\
  \citenamefont {Fang}}]{SSG3}%
  \BibitemOpen
  \bibfield  {author} {\bibinfo {author} {\bibfnamefont {Y.}~\bibnamefont
  {Jiang}}, \bibinfo {author} {\bibfnamefont {Z.}~\bibnamefont {Song}},
  \bibinfo {author} {\bibfnamefont {T.}~\bibnamefont {Zhu}}, \bibinfo {author}
  {\bibfnamefont {Z.}~\bibnamefont {Fang}}, \bibinfo {author} {\bibfnamefont
  {H.}~\bibnamefont {Weng}}, \bibinfo {author} {\bibfnamefont {Z.-X.}\
  \bibnamefont {Liu}}, \bibinfo {author} {\bibfnamefont {J.}~\bibnamefont
  {Yang}},\ and\ \bibinfo {author} {\bibfnamefont {C.}~\bibnamefont {Fang}},\
  }\bibfield  {title} {\bibinfo {title} {Enumeration of spin-space groups:
  Toward a complete description of symmetries of magnetic orders},\ }\href
  {https://doi.org/10.1103/PhysRevX.14.031039} {\bibfield  {journal} {\bibinfo
  {journal} {Phys. Rev. X}\ }\textbf {\bibinfo {volume} {14}},\ \bibinfo
  {pages} {031039} (\bibinfo {year} {2024})}\BibitemShut {NoStop}%
\bibitem [{\citenamefont {Zhang}\ and\ \citenamefont
  {Xiong}(2012)}]{ferro_symm}%
  \BibitemOpen
  \bibfield  {author} {\bibinfo {author} {\bibfnamefont {W.}~\bibnamefont
  {Zhang}}\ and\ \bibinfo {author} {\bibfnamefont {R.-G.}\ \bibnamefont
  {Xiong}},\ }\bibfield  {title} {\bibinfo {title} {Ferroelectric
  metal--organic frameworks},\ }\href {https://doi.org/10.1021/cr200174w}
  {\bibfield  {journal} {\bibinfo  {journal} {Chem. Rev.}\ }\textbf {\bibinfo
  {volume} {112}},\ \bibinfo {pages} {1163} (\bibinfo {year}
  {2012})}\BibitemShut {NoStop}%
\end{thebibliography}
%

\clearpage
\section*{Supplementary information}

\subsection{Many-body chiral symmetry in our model}
\label{MBCS}

We introduce here the many-body chiral symmetry relevant to the interacting fermion model given in Eq.~\eqref{model} of the main text.
Following Refs.~\cite{Gurarie2011, Kimura2019}, the many-body chiral symmetry $\hat{U}_{\Gamma}$ is defined by the relation
\begin{eqnarray}
\hat{U}_{\Gamma}^{\dagger} \, \hat{H}^{*} \, \hat{U}_{\Gamma} = \hat{H}, 
\label{eq:manybody_chiral_def}
\end{eqnarray}
where $\hat{H}$ is the many-body Hamiltonian and $\hat{U}_{\Gamma}$ is a unitary chiral operator with $\hat{U}_{\Gamma}^2 = 1$.
The action of $\hat{U}_{\Gamma}$ on the fermionic creation and annihilation operators is given by
$\hat{U}_{\Gamma}^{\dagger} \, \hat{c}_{\ve{i}n}^{\dagger} \, \hat{U}_{\Gamma}= \sum_m \hat{c}_{\ve{i}m} \, (U_{\Gamma}^{\dagger})_{mn}$ and 
$\hat{U}_{\Gamma}^{\dagger} \, \hat{c}_{\ve{i}n} \, \hat{U}_{\Gamma}= \sum_m (U_{\Gamma})_{nm} \, \hat{c}_{\ve{i}m}^{\dagger}$,
where $\ve{i}$ labels lattice sites and $n$ denotes internal degrees of freedom such as sublattice and spin.
Furthermore, at the single-particle level, this operator acts on the quadratic block of the Hamiltonian $h$ by anticommuting with it:
$\hat{U}_{\Gamma}^{\dagger} h \, \hat{U}_{\Gamma} = -h$. 

In our model, we define the chiral symmetry operator as
\begin{eqnarray}
\hat{U}_{\Gamma} = \prod_{\ve{j}} \left[
  e^{i\pi \hat{n}_{\ve{j}\downarrow}} \,
  \exp\left(
    i \frac{\pi}{2} \, \text{sgn}(\nu_{\ve{j}})
    \left( \hat{c}_{\ve{j}\uparrow} \hat{c}_{\ve{j}\downarrow}
         + \hat{c}_{\ve{j}\downarrow}^\dagger \hat{c}_{\ve{j}\uparrow}^\dagger \right)
  \right)
\right],\nonumber \\
\label{eq:chiral_operator}
\end{eqnarray}
where $\text{sgn}(\nu_{\ve{j}}) = +1$ for sublattices $A,D$ and $-1$ for $B,C$.
The operator $\hat{U}_{\Gamma} $ acts locally on each site and exchanges particles and holes, flips the spin, and applies a sublattice-dependent sign factor. 
It satisfies the transformation rules:
\begin{eqnarray}
&& \hat{U}_{\Gamma}^{\dagger} \, \hat{c}_{\ve{j}\uparrow} \, \hat{U}_{\Gamma} = \text{sgn}(\nu_{\ve{j}}) \, \hat{c}_{\ve{j}\downarrow}^{\dagger}, \nonumber \\
&&\hat{U}_{\Gamma}^{\dagger} \, \hat{c}_{\ve{j}\downarrow} \, \hat{U}_{\Gamma}= -\text{sgn}(\nu_{\ve{j}}) \, \hat{c}_{\ve{j}\uparrow}^{\dagger}.
\end{eqnarray}

We now examine the single-particle Hamiltonian defined in Eq.~\eqref{model} of the main text.
Written in the $8\times8$ basis (sublattice $\times$ spin), the Hamiltonian takes the form $h_8(\mathbf{k}) = h_4(\mathbf{k}) \otimes \sigma_0$, where $ h_4(\mathbf{k}) $ acts on the sublattice space and $\sigma_0 $ is the $2\times2$ identity in spin space.
We define the chiral matrix as
\begin{eqnarray}
\Gamma_8 = \mathrm{diag}(-1,+1,+1,-1)\otimes i\sigma_y,
\end{eqnarray}
where $\mathrm{diag}(-1,+1,+1,-1)$ acts on the four-sublattice structure (A,B,C,D), $\mu_0$ is the identity on the A/D vs B/C subspace, and $ \sigma_y$ is the Pauli matrix acting on spin.
We confirm that $\{\Gamma_8, h_8(\mathbf{k}) \} = 0$, demonstrating the chiral symmetry under the transformation.

The hopping part $\hat{H}_t$ remains invariant under the chiral transformation, $\hat{U}_{\Gamma}^\dagger \hat{H}_t \hat{U}_{\Gamma} = \hat{H}_t$, as the two minus signs -- one from the matrix conjugation and one from fermionic anticommutation -- exactly cancel.
The interaction term,
$\hat{H}_U = U \sum_{\ve{j}} \left( \hat{n}_{\ve{j}\uparrow} - \tfrac{1}{2} \right) \left( \hat{n}_{\ve{j}\downarrow} - \tfrac{1}{2} \right)$,
is also invariant under $\hat{U}_{\Gamma}$, since
\begin{eqnarray}
&&\hat{U}_{\Gamma}^\dagger \hat{n}_{\ve{j}\uparrow} \hat{U}_{\Gamma} = 1 - \hat{n}_{\ve{j}\downarrow}, \nonumber \\
&&\hat{U}_{\Gamma}^\dagger \hat{n}_{\ve{j}\downarrow} \hat{U}_{\Gamma}= 1 - \hat{n}_{\ve{j}\uparrow}.
\end{eqnarray}
This implies $\hat{U}_{\Gamma}^\dagger \hat{H}_U \hat{U}_{\Gamma} = \hat{H}_U$, and hence the full many-body Hamiltonian $\hat{H}= \hat{H}_t+\hat{H}_U$ [Eq.~\eqref{model} of the main text] commutes with the chiral operator:
\begin{eqnarray}
\hat{U}_{\Gamma}^\dagger \hat{H} \hat{U}_{\Gamma} = \hat{H}.
\end{eqnarray}

We also verify that the three components of the antiferromagnetic (AFM) N\'eel order parameter,
\begin{eqnarray}
\mathbf{N} = \sum_{\ve{j}} \text{sgn}(\nu_{\ve{j}}) \, \hat{\mathbf{S}}_{\ve{j}},
\end{eqnarray}
are invariant under the transformation by $\hat{U}_{\Gamma}$. 
Each spin component $\hat{S}_{\ve{j}}^\alpha$ ($\alpha = x, y, z$) transforms into itself, and therefore the AFM order vector $\mathbf{N}$ is even under the chiral transformation.

To summarize, the many-body chiral symmetry defined above remains unbroken in the altermagnetic insulating state that emerges due to the Hubbard interaction in the Hamiltonian Eq.~\eqref{model} of the main text.
It is important to note that the alternating on-site potential term, $\hat{H}_\eta = \sum_{\ve{j}} \eta_{\ve{j}} \left(\hat{n}_{\ve{j}\uparrow}+\hat{n}_{\ve{j}\downarrow} \right) $, which is introduced as an additional term to Eq.~\eqref{model}, explicitly breaks the chiral symmetry.
Indeed, under the chiral transformation, this term behaves as
\begin{eqnarray}
\hat{U}_{\Gamma}^\dagger \hat{H}_\eta \hat{U}_{\Gamma} &&= \sum_{\ve{j}} \eta_{\ve{j}} \left(2 - \hat{n}_{\ve{j}\downarrow}-\hat{n}_{\ve{j}\uparrow} \right) \nonumber \\
&&\neq \hat{H}_\eta,
\end{eqnarray}
which explicitly shows that the $\eta$-term does not commute with the chiral operator and therefore breaks the chiral symmetry.

\subsection{Quantum Monte Carlo determination of altermagnetic order}
\label{QMC}
To establish the presence and nature of altermagnetic order in the interacting model introduced in the main text, we perform auxiliary-field quantum Monte Carlo simulations on torus geometries.
The results are obtained on lattices with $L \times L$ unit cells ($4L^2$ sites) and periodic boundary conditions.
We compute the equal-time correlation functions of the fermion spin, $\hat{\ve{ S}}_{\ve{i}}= \frac{1}{2}\sum_{s,s'} \hat{c}_{\ve{i}s}^{\dagger}\ve{\sigma}_{s,s'} \hat{c}^{}_{\ve{i}s'}$, where $\ve{\sigma}$ corresponds to the vector of Pauli spin-1/2 matrices.
Due to the larger unit cell, these correlation functions are $4\times 4$ matrices of the form $C^S_{\ve{R}\gamma,\ve{R}'\delta}=\langle (\hat{{\ve{S}}}_{\ve{R}\gamma}-\langle \hat{{\ve{S}}}_{\ve{R}\gamma} \rangle)\cdot(\hat{{\ve{S}}}_{\ve{R}'\delta}-\langle \hat{{\ve{S}}}_{\ve{R}'\delta} \rangle)\rangle$ where $\ve{R}, \ve{R}'$ label the unit cell and $\gamma,\delta$ the orbitals.
After diagonalizing the corresponding structure factors $C^S_{\gamma \delta}(\ve{q})=\frac{1}{L^2}\sum_{\ve{R}\ve{R}'}C^S_{\ve{R}\gamma,\ve{R}'\delta}e^{i {\ve q}\cdot (\ve{R}-\ve{R}')}$,
we calculated the renormalization-group invariant correlation ratio \cite{Binder1981, Pujari16}
\begin{eqnarray}
R^{S}=1-\frac{\lambda_1({\ve q}_0+\delta {\ve q})}{\lambda_1({\ve q}_0)}
\label{eq:CR}
\end{eqnarray}
using the largest eigenvalue $\lambda_1({\ve q})$; ${\ve q}_0$ is the ordering wave vector, ${\ve q}_0 + \delta {\ve q}$ a neighboring wave vector.
A long-range AFM order implies a divergence of the corresponding $\lambda_1({\ve q}_0 = \Gamma)$.
Thus, $R^S\to 1$ for $L\to\infty$ in the corresponding ordered state, whereas $R^S\to 0$ in the disordered state.
At the critical point, $R^S$ is scale-invariant for sufficiently large $L$, leading to a crossing of results for different $L$. 

\begin{figure}[t]
\centering
\centerline{\includegraphics[width=0.5\textwidth]{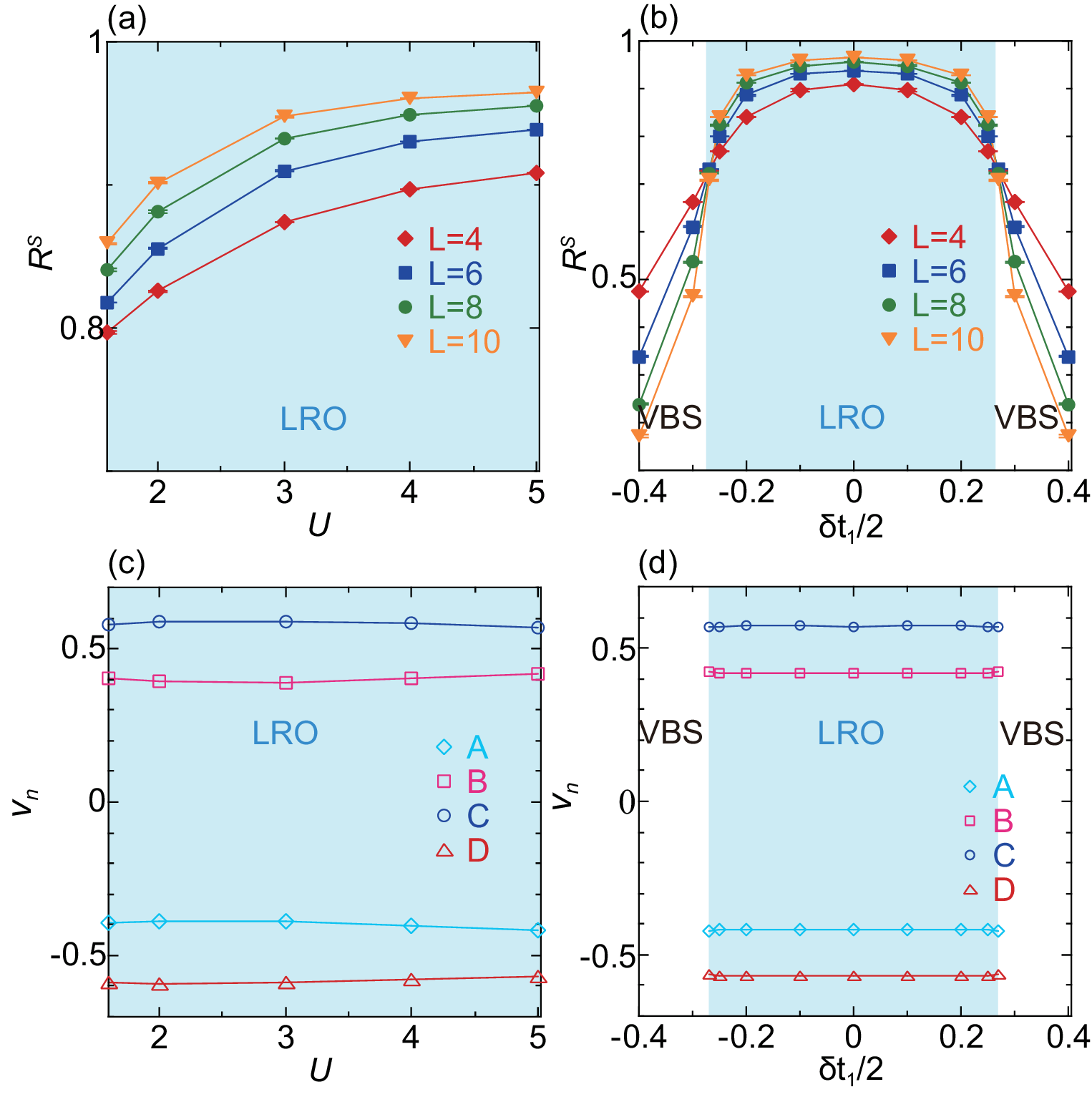}}
\caption{\label{fig:CR}
Correlation ratio $R^{S}$ for the AFM order as a function of $U$ [(a), with $\delta t_1=0$] and $\delta t_1$ [(b), with $U=5$] for different lattice sizes $L$.
In the shaded regions, $R^{S}$ increases with increasing $L$, indicating the presence of long-range antiferromagnetic order (LRO).
Outside these regions, $R^{S}$ decreases with $L$, signaling the absence of long-range antiferromagnetic order and a transition to a valence-bond solid (VBS) phase.
Panels (c) and (d) show the corresponding eigenvector $\ve{v}=(v_\text{A},v_\text{B},v_\text{C},v_\text{D})$ of the largest eigenvalue $\lambda_1({\ve q_0})$, characterizing the spin structure of the LRO phase for the same parameters as in panels (a) and (b), respectively.
The lattice size is $L=10$ for panels (c) and (d).
}
\end{figure}

We find that the system hosts an altermagnetic Mott-insulating phase for a range of parameters $U$ and $\delta t_1$, as shown in Fig.~\ref{fig:CR}. 
Figure~\ref{fig:CR}(a) shows the results as a function of $U$ at $\delta t_1=0$.
The onset of the long-range spin order is signaled by an increase in $R^{S}$ with increasing $L$.
Within the range of parameters we investigated, we confirm the presence of long-range spin order. 
Results as a function of $\delta t_1$ for $U=5$ are presented in Fig.~\ref{fig:CR}(b), showing that the antiferromagnetic order persists for a finite range of $\delta t_1$.

To characterize the nature of the magnetic order in the long-range ordered phase, we analyze the eigenvector $v_l$ corresponding to the largest eigenvalue $\lambda_1({\ve q_0})$.
Here the index $l$ runs over sublattices A, B, C, and D.
Figures~\ref{fig:CR}(c) and (d) show the evolution of the eigenvector components $(v_\text{A},v_\text{B},v_\text{C},v_\text{D})$ as a function of $U$ and $\delta t_1$, respectively.
In the shaded regions where the correlation ratio $R^{S}$ indicates the presence of long-range antiferromagnetic order, the eigenvector exhibits a consistent pattern: $(v_\text{A},v_\text{B},v_\text{C},v_\text{D})=(-m,m,m',-m')$, with $\sum_l v_{l}= 0 $, consistent with a fully compensated collinear antiferromagnetic state.
In this ordered state, distinct magnetic sublattices are not connected by translation or inversion combined with time reversal, which is consistent with the defining symmetry of an altermagnet.
A key feature of the altermagnetic Mott-insulating state that we observe is that, for any values of $\delta t_1$, it spontaneously breaks both time-reversal symmetry $\mathcal{T}$ and glide symmetry $\mathcal{G}_{x-y}$, while preserving their combination, $\mathcal{G}_{x-y}\mathcal{T}$.
As $|\delta t_1|$ is further increased, the correlation ratio provides no clear evidence for the AFM order, indicating a transition to a valence-bond solid (VBS) phase; accordingly, the transition is most probably in the 3D O(3) universality class.

\begin{figure}
\centering
\centerline{\includegraphics[width=0.5\textwidth]{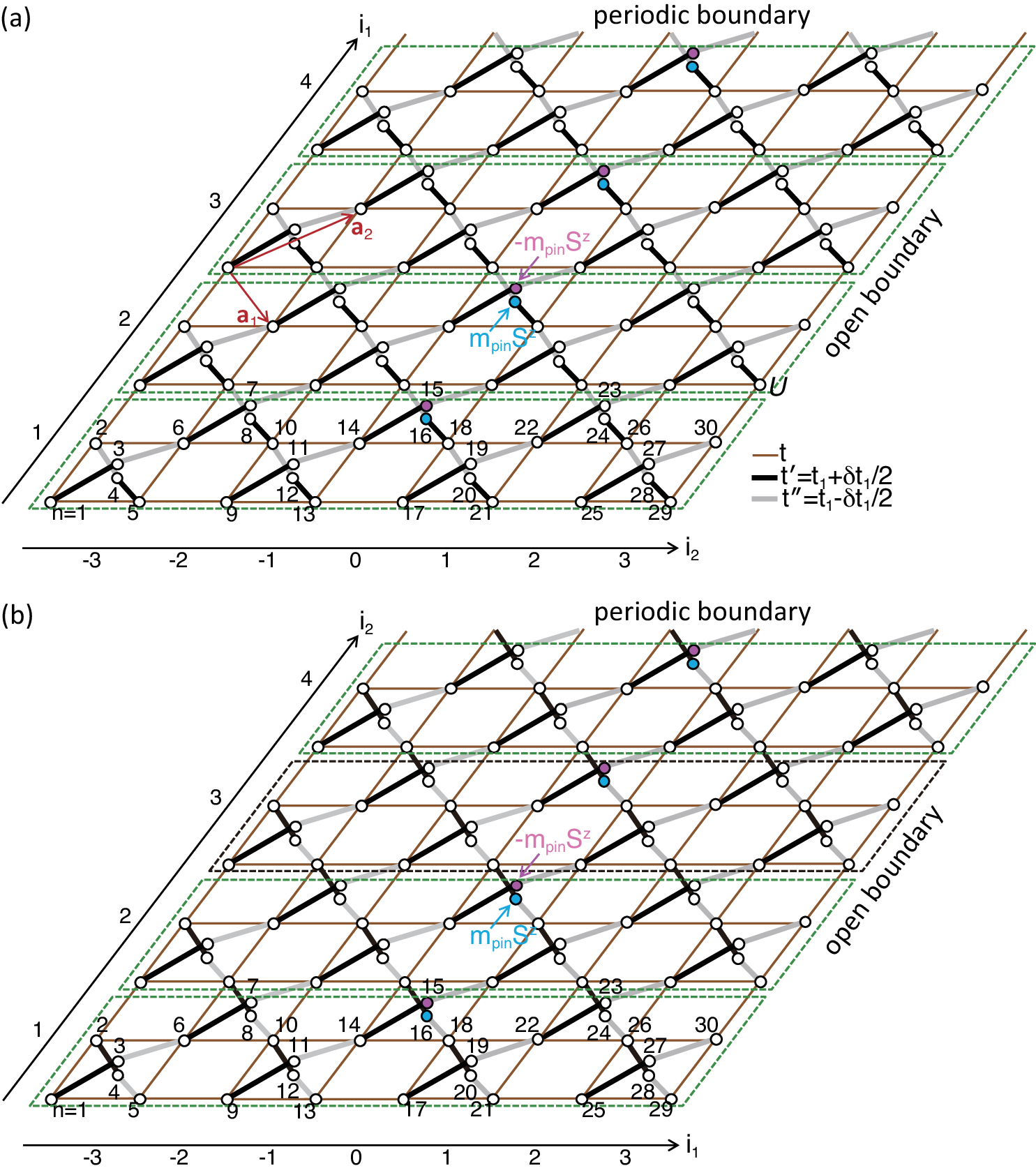}}
\caption{\label{fig:model-OB}
Cylindrical lattice geometries used in the QMC simulations.
(a) Geometry with periodic boundaries along $\mathbf{a}_2-\mathbf{a}_1$ and open boundaries along $\mathbf{a}_2+\mathbf{a}_1$.
The system is constructed with $L$ unit cells stacked along the periodic boundary direction, labeled by $i_1$.
Each unit cell extends in the open boundary direction and contains $N_{\rm orb}$ orbitals, indexed by $i_2$.
Pinning fields $m_{\rm pin}\hat{S}^z$ are applied at the central layer ($i_2=0$).
As an example, we show a lattice with $L=4$ unit cells and $N_{\rm orb}=30$ orbitals.
(b) Geometry obtained by rotating the system by 90 degrees, exchanging the roles of the two directions.
}
\end{figure}

\subsection{Cylindrical lattice setups for QMC simulations }
\label{QMC-OB}
In the main text, we have analyzed spin and charge polarization using cylindrical geometries in our quantum Monte Carlo simulations with periodic boundary conditions along one direction and open boundary conditions along the orthogonal direction.
For clarity, we summarize here the specific lattice setups employed in our simulations and illustrate them in Fig.~\ref{fig:model-OB}.
Figure~\ref{fig:model-OB}(a) shows the geometry used in the main text.
We impose periodic boundary conditions along the $\mathbf{a}_2-\mathbf{a}_1$ (i.e., $x$-$y$) direction and open boundary conditions along the $\mathbf{a}_2+\mathbf{a}_1$ (i.e., $x$+$y$) direction.
The lattice is constructed with $L$ unit cells along the periodic $\mathbf{a}_2-\mathbf{a}_1$ direction, labeled by the index $i_1$.
Each unit cell contains $N_{\rm orb}$ orbitals aligned along the open $\mathbf{a}_2+\mathbf{a}_1$ direction, indexed by $i_2$.
We apply a weak pinning field $\pm m_{\rm pin} \hat{S}^z_{i_1,i_2=0}$~\cite{Assaad13} at the central layer (for which $i_2=0$) in the periodic direction on the D and C sublattice respectively.
As an illustration, Fig.~\ref{fig:model-OB}(a) shows a finite lattice with $L=4$ unit cells and $N_{\rm orb}=30$ orbitals.
Figure~\ref{fig:model-OB}(b) shows the geometry used in the main text, obtained from the setup in Fig.~\ref{fig:model-OB}(a) by rotating the system by 90 degrees, thereby exchanging the roles of the $\mathbf{a}_2-\mathbf{a}_1$ and $\mathbf{a}_2+\mathbf{a}_1$ axes.

The spin accumulation exhibits an exponentially decaying envelope, $m^z(d) = e^{-d/\xi} f(d)$, where $d$ denotes the distance from the edge and $\xi$ represents the characteristic length scale. 
The weak pinning field explicitly breaks the SU(2) spin symmetry down to U(1) corresponding to spin rotations around the $z$-axis. 
Thereby, fluctuations of the $z$-component of spin correspond to amplitude fluctuations, the Higgs mode \cite{Lohofer15}. 
This bulk mode being gapped is consistent with a spin accumulation at the edges of the system.

\subsection{Finite-size analysis of ferro-spinetic polarization}
\label{Ps-FSS}

\begin{figure}[t]
\centering
\centerline{\includegraphics[width=0.5\textwidth]{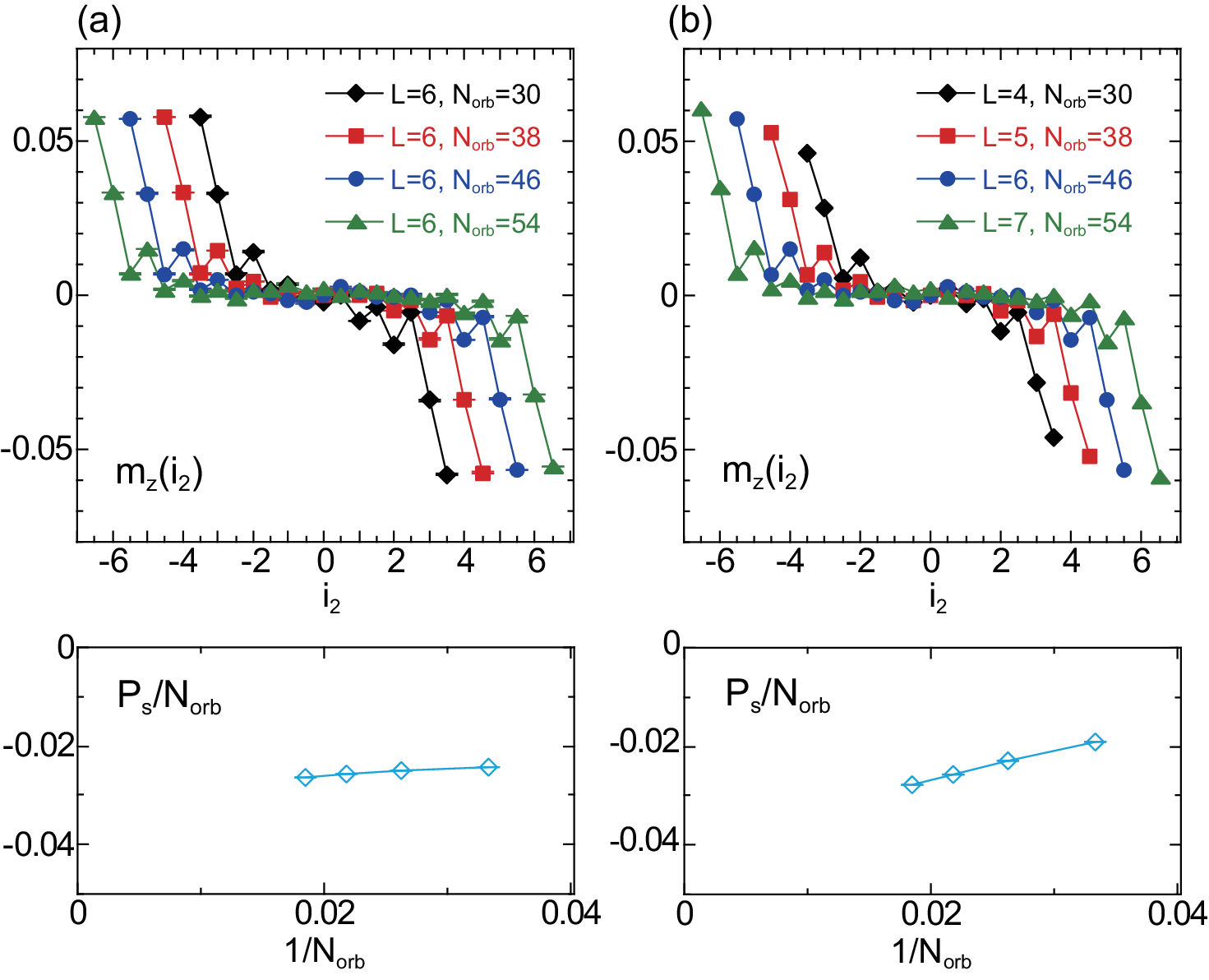}}
\caption{\label{fig:Polarization-FSS}
Real-space distribution of magnetization $m_z(i)$.
From these quantities, the polarizations are evaluated as $P_s=\sum_{i} i m_z(i)$.
The setup is the same as in Fig.~2(a) of the main text  (see also Fig.~\ref{fig:model-OB}(a)).
Panel (a) shows results for systems with a fixed number of unit cells along the periodic ($i_1$) direction, $L=6$, while the number of orbitals along the open ($i_2$) direction, $N_{\rm orb}$, is increased to probe edge accumulation.
Panel (b) shows results where both the periodic ($i_1$) and open ($i_2$) directions are increased simultaneously, keeping the overall shape fixed to reach the thermodynamic limit.
Here, $U = 5$,  $\delta t_1/2=0.2$, and $m_{\rm pin} = 0.01$.
}
\end{figure}

To clarify the behavior of the ferro-spinetic polarization in the thermodynamic limit, we perform a finite-size analysis. 
Figure~\ref{fig:Polarization-FSS} illustrates the system size dependence of the real-space distribution of magnetization $m_z(i)$, and the polarizations evaluated as $P_s=\sum_{i} i m^z(i)$, calculated under the same setup as in the main text Fig.~\ref{fig:Polarization}(a).

Figure~\ref{fig:Polarization-FSS}(a) shows the results with a fixed system length along the periodic boundary direction and increasing system size along the open boundary direction to examine the edge accumulation behavior. 
As the system is extended in the open direction, the spin accumulation remains confined near the edges, and the bulk region stays unpolarized.
Figure~\ref{fig:Polarization-FSS}(b) presents the scaling where both the periodic and open directions are enlarged proportionally, maintaining a fixed aspect ratio to access the thermodynamic limit. 
The magnetization profiles $m^z(i)$ suggest that the spin accumulation near the edges becomes more pronounced with increasing system size. 
The data for $P_s$ suggest that it tends toward a finite value in the thermodynamic limit.

\subsection{Electronic properties of Mn-MOF} 

\begin{figure}
\centering
\centerline{\includegraphics[width=0.45\textwidth]{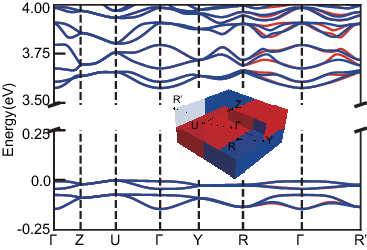}}
    \caption{The band structure of Mn-MOF. The schematic $d-$wave distribution is shown in the insert with high symmetry points are denoted. Opposite orderings of spin-splits are along $R-\Gamma-R'$.}
    \label{fig:Mn-MOF_band}
\end{figure}

As shown in Fig.~\ref{fig:Mn-MOF_band}, Mn-MOF is an insulator with a relatively large band gap of 3.59 eV. 
Spin splittings occur in regions away from high-symmetry points and lines, as indicated in the Brillouin zone inset of Fig.~\ref{fig:Mn-MOF_band}.
In addition to the glide-mirror symmetry that connects the opposite magnetic sublattices, the screw-rotation symmetry also acts as a generator and is expressed as $\{-1||2_z|\bm{t}'\}$.
These two generators give rise to a $d-$wave spin polarized Fermi surface exhibiting a planar character within the $k_y-k_z$ plane.

\subsection{General spin symmetry constraints on ferro-spinetic altermagnetism}
Here we focus only on collinear altermagnets, for which $\{TU_{\bm{n}}(\pi)||E|0\}$ is the general symmetry constraint in the spin space group~\cite{SSG1,SSG2,SSG3}. 
In other words, the $SU(2)$ symmetry is preserved in our many-body model rendering the spin polarization $\bm{P}_s$ a vector pointing only in real space. 
In the following derivation, we define the spin polarization further as $\bm{P}_s = \bm{P}^{\uparrow}_s - \bm{P}^{\downarrow}_s$, and apply the constraints phenomenologically.

First, we state the general condition for obtaining a non-zero $\bm{P}_s$. 
As a first step, we confirm that if both inversion ($I$) and time-reversal ($T$) symmetries are present, they guarantee $\bm{P}_s = 0$.
We define the spin polarization as
\begin{eqnarray}
\bm{P}_s  & = \int d^3\bm{r} [P^{\uparrow}(\bm{r}) - P^{\downarrow}(\bm{r})].
\end{eqnarray}
Under time reversal ($T$), one has $TP^{\uparrow/\downarrow}(\bm{r})=P^{\downarrow/\uparrow}(\bm{r})$, which yields
\begin{eqnarray}
\bm{P}_s &&= T\bm{P}_s  = \int d^3\bm{r} [TP^{\uparrow}(\bm{r}) - TP^{\downarrow}(\bm{r})] \nonumber \\
&&= \int d^3\bm{r} [P^{\downarrow}(\bm{r}) - P^{\uparrow}(\bm{r})] =  -\bm{P}_s = 0. \nonumber
\end{eqnarray}
Similarly, under inversion ($I$), $IP^{\uparrow/\downarrow}(\bm{r})=P^{\uparrow/\downarrow}(-\bm{r})$, so that
\begin{eqnarray}
\bm{P}_s  & &= I\bm{P}_s = \int d^3\bm{r} [IP^{\uparrow}(\bm{r}) - IP^{\downarrow}(\bm{r})] \nonumber \\
&&= \int d^3\bm{r} [P^{\uparrow}(-\bm{r}) - P^{\downarrow}(-\bm{r})]
= -\bm{P}_s= 0. \nonumber
\end{eqnarray}

In the following, we will only discuss and derive the constraints from the altermagnetic symmetries that connect the spin-up and spin-down sublattices in real space.
Here $g$ denotes the generator of the symmetry operation, $n$ indicates the rotational order ($n=2,4,6$), and $D(g)$ is the corresponding matrix representation acting on $\bm{P}_s$.
For the magnetic sublattice site symmetries, we argue that these can act on charge polarization but not spin polarization, which follows the same rule as in ferroelectricity \cite{ferro_symm}. 
If we have the generator $g=\{-1||n_{\bm{m}}\}$ with $n=2,4,6$, then
$g_oP^{\uparrow/\downarrow}(\bm{r})  = P^{\downarrow/\uparrow}(g_o\bm{r})$, $g_o = g^{2i+1}$, and
$g_eP^{\uparrow/\downarrow}(\bm{r})  = P^{\uparrow/\downarrow}(g_e\bm{r})$, $g_e = g^{2i}$, $i\in \mathbb{N}$.
\begin{eqnarray}
\bm{P}_s  = &&g_o\bm{P}_s= \int d^3\bm{r} [P^{\downarrow}(g_o\bm{r}) - P^{\uparrow}(g_o\bm{r})]\nonumber \\
= &&\int d^3\bm{r}_{\parallel} [P^{\downarrow}(\bm{r}_{\parallel}) - P^{\uparrow}(\bm{r}_{\parallel})] \nonumber \\
    &&+ \int d^3\bm{r}_{\perp} [P^{\downarrow}(g_o\bm{r}_{\perp}) - P^{\uparrow}(g_o\bm{r}_{\perp})] \nonumber \\
= &&-\bm{P}_{s,\parallel} + D(g_o^{-1})\int d^3g_o\bm{r}_{\perp} [P^{\downarrow}(g_o\bm{r}_{\perp}) - P^{\uparrow}(g_o\bm{r}_{\perp})]\nonumber \\
= &&-\bm{P}_{s,\parallel} - D(g_o^{-1})\bm{P}_{s,\perp}, \nonumber \\
\bm{P}_s  = &&g_e\bm{P}_s  =\int d^3\bm{r} [P^{\uparrow}(g_e\bm{r}) - P^{\downarrow}(g_e\bm{r})]\nonumber \\
= &&\int d^3\bm{r}_{\parallel} [P^{\uparrow}(\bm{r}_{\parallel}) - P^{\downarrow}(\bm{r}_{\parallel})] \nonumber \\
&&+ \int d^3\bm{r}_{\perp} [P^{\uparrow}(g_e\bm{r}_{\perp}) - P^{\downarrow}(g_e\bm{r}_{\perp})] \nonumber \\
 = &&\bm{P}_{s,\parallel} + D(g_e^{-1})\int d^3g_e\bm{r}_{\perp} [P^{\uparrow}(g_e\bm{r}_{\perp}) - P^{\downarrow}(g_e\bm{r}_{\perp})]\nonumber \\
 =&& \bm{P}_{s,\parallel} + D(g_e^{-1})\bm{P}_{s,\perp}. \nonumber
\end{eqnarray}
For the improper rotational symmetry $g=\{-1||-n_{\bm{m}}\}$ with $n=2,4,6$,
\begin{eqnarray}
\bm{P}_s   &&= g_o\bm{P}_s=\bm{P}_{s,\parallel} - D(g_o^{-1})\bm{P}_{s,\perp},\nonumber \\
\bm{P}_s   &&= g_e\bm{P}_s = \bm{P}_{s,\parallel} + D(g_e^{-1})\bm{P}_{s,\perp}. \nonumber
\end{eqnarray}

We now derive the crystal symmetry constraints explicitly for the cases $n=2$, $n=4$, and $n=6$.
For $n=2$, the representation is  $D(2^{-1}) = -1_{2\times 2}$, which yields
\begin{eqnarray}
\bm{P}_s  =  -\bm{P}_{s,\parallel} + \bm{P}_{s,\perp},~~~\therefore \bm{P}_{s,\parallel} = 0,~\bm{P}_{s,\perp} \neq 0. \nonumber
\end{eqnarray}
For the improper $n=2$ rotation, $D(-2^{-1}) = 1_{2\times 2}$, leading to
\begin{eqnarray}
\bm{P}_s  =  \bm{P}_{s,\parallel} - \bm{P}_{s,\perp},~~~\therefore \bm{P}_{s,\parallel} \neq 0,~\bm{P}_{s,\perp}= 0. \nonumber
\end{eqnarray}

For $n=4$, $D(4^{-1}) = \begin{bmatrix} 0 & 1 \\ -1 & 0 \end{bmatrix}$, $D(2) = \begin{bmatrix} -1 & 0 \\ 0 & -1 \end{bmatrix}$.
Hence,
\begin{eqnarray}
&&\bm{P}_s  =  -\bm{P}_{s,\parallel} - D(4^{-1})\bm{P}_{s,\perp} =  \bm{P}_{s,\parallel} + D(2)\bm{P}_{s,\perp}, \nonumber \\
\therefore~&& \bm{P}_{s,\parallel} = 0,~\bm{P}_{s,\perp} = 0. \nonumber
\end{eqnarray}
For the improper $n=4$ rotation, $ D(-4^{-1}) = \begin{bmatrix} 0 & -1 \\ 1 & 0 \end{bmatrix} $, which gives
\begin{eqnarray}
&&\bm{P}_s  =  \bm{P}_{s,\parallel} - D(-4^{-1})\bm{P}_{s,\perp} =  \bm{P}_{s,\parallel} + D(2)\bm{P}_{s,\perp}, \nonumber \\
\therefore~&&\bm{P}_{s,\parallel} \neq 0,~\bm{P}_{s,\perp} = 0. \nonumber 
\end{eqnarray}

For $n=6$, $D(6^{-1}) = \begin{bmatrix} \frac{1}{2} & \frac{\sqrt{3}}{2} \\ -\frac{\sqrt{3}}{2} & \frac{1}{2} \end{bmatrix}$ so that
\begin{eqnarray}
\bm{P}_s  =  -\bm{P}_{s,\parallel} - D(6^{-1})\bm{P}_{s,\perp},~~~\therefore \bm{P}_{s,\parallel} = 0,~\bm{P}_{s,\perp} = 0. \nonumber
\end{eqnarray}
For the improper $n=6$ rotation, $D(-6^{-1}) = \begin{bmatrix} -\frac{1}{2} & -\frac{\sqrt{3}}{2} \\ \frac{\sqrt{3}}{2} & -\frac{1}{2} \end{bmatrix}$, which gives
\begin{eqnarray}
\bm{P}_s  =  \bm{P}_{s,\parallel} - D(-6^{-1})\bm{P}_{s,\perp},~~~\therefore \bm{P}_{s,\parallel} \neq 0,~\bm{P}_{s,\perp} = 0. \nonumber
\end{eqnarray}
Finally, the results are summarized in Table~\ref{tab:crystal_spin}.

\begin{table}[t]
    \centering
    \begin{tabular}{|c|c|c|c|c|c|c|}
          \hline
          & 2 & -2 (m) & 4 & -4 & 6 & -6 \\
          \hline
        $\bm{P}_{s,\parallel}$ & $\times$ & $\checkmark$ & $\times$ & $\checkmark$ & $\times$ & $\checkmark$ \\
          \hline
         $\bm{P}_{s,\perp}$ & $\checkmark$ & $\times$ & $\times$ & $\times$ &  $\times$ &  $\times$ 
         \\
          \hline
    \end{tabular}
    \caption{\label{tab:crystal_spin}
    The spin polarization orientations under different rotational symmetry constraints.}
\end{table}

\end{document}